\documentclass[11pt]{article}

\usepackage{fullpage,natbib,amssymb,amsfonts,amsmath,graphicx,xcolor,mathtools,pdfpages}
\usepackage{nicematrix,colortbl,soul}

\def\pconv{\smash{\mathop{\longrightarrow}\limits^p}}     
\def\vec{\text{vec}}

\begin{document}

\title{Constructing  High Frequency   Economic Indicators by Imputation}

\author{Serena Ng\footnote{Corresponding author: Department of Economics, Columbia University and NBER. Email: serena.ng at columbia.edu 420 W. 118 St. MC 3308, New York, NY 10027}
\and Susannah Scanlan\footnote{Department of Economics, Columbia University. \newline 
We thank Siem Koopman, Jushan Bai, Joerg  Breitung, Krishna Kamepalli, and Siem Koopmans for many helpful discussions and  comments. This research is supported by the National Science Foundation (SES 2018369)} 
}
\date{\today}
\maketitle
\setcounter{page}{0}

\begin{abstract}
Monthly and weekly economic indicators are often taken to be the largest common factor estimated from  high and low frequency data,  either separately or jointly.  To incorporate mixed frequency information  without directly modeling them,  we target a low frequency diffusion index that is already available, and treat high frequency values as missing. We impute these values using multiple factors estimated from the high frequency data. In the empirical examples considered, static matrix completion that does not account for serial correlation in the idiosyncratic errors yields imprecise estimates of the missing values  irrespective of how the factors are estimated.  Single equation and  systems-based dynamic procedures that account for serial correlation yield  imputed  values that are closer to the observed low frequency ones. This is the case in the counterfactual exercise that imputes the monthly values of  consumer sentiment series before 1978 when the data was released only on a quarterly basis.  This is also the case for  a weekly version of the  CFNAI index of economic activity that is imputed using seasonally unadjusted data. The  imputed series reveals episodes of increased variability of weekly economic information that are masked by the monthly data, notably around the 2014-15 collapse in oil prices.



\end{abstract}

\section{Introduction}

Traditional economic time series are most frequently provided by government agencies on a  monthly or quarterly  basis, and it is not surprising that empirical work is predominantly  conducted at these sampling frequencies. But increasingly available are weekly and daily data collected officially or privately over  shorter spans, possibly irregularly spaced. This creates  opportunities to explore macroeconomic relations that were previously studied at conventional lower frequencies.


Previous work on harnessing the informational content of weekly macroeconomic variables falls into two broad categories. The first category, as exemplified by the weekly economic index (WEI) of \citet{stock:13}, only uses high frequency data. 
The WEI, constructed  as  the first principal component of eight weekly series, proved to be useful during  the  government shutdown of 2013 when the release of official data was delayed. Using the same methodology, \citet{lewis-mertens-stock} expanded the panel of data for monitoring economic activities after the outbreak of COVID-19 in early 2020. Though the WEI is calibrated to quarterly GDP growth,  information in  monthly or quarterly data is not used.

In the second category, researchers jointly model data at both high and low frequencies in a dynamic factor model such as in \citet{aruoba-diebold-scotti} (ADS index)\footnote{\citet{cglms:20}, \citet{primiceri-tambalotti:20}, \citet{moran-etal:20}, \citet{lewis-mertens-stock}, \citet{fms:20}, \citet{hkops:20}, \citet{cglms:20},  \citet{adp:21}, \citet{galvao-owyang:20}, \citet{adp:21} among many others.}.  \citet{baumeister-leiva-sims:21}, for example, estimate a weekly dynamic  factor model   that  allows  each state to contribute to a national economic condition indicator (ECI)  that is  a weighted sum of weekly, monthly, and quarterly data using the factor loadings as weights. Fully specifying the model is useful in many applications, but as is the case with systems estimation,  internal coherence comes with the cost that   misspecification  in one part of the system can have system wide effects.  Furthermore, joint estimation of factors from  different sampling frequencies necessarily enlarges the state space, and imposing the stock-flow constraints along the lines of \citet{mariano-murasawa:03} can be computationally burdensome.

In this paper, we treat  the target low frequency series $\{Y_{L,t}\}$, $t=1,\ldots, T_o$  as an incompletely observed  high frequency series  $\{Y_{H,t+j/m_t}\}, j=1,\ldots,m_t$  of length $T=\sum_{t=1}^{T_o} m_t$.   The series has missing values  because  it is observed  only  once in the $m_t$ sub-periods between $t$ and $t+1$.  Here, (L, H) can be (yearly, quarterly) with $m=4$, or (quarterly, monthly) with $m=3$. More challenging is the (monthly, weekly) case when $m_t$ varies with month $t$.  The problem of estimating the high frequency indicator is then turned into one of estimating the missing  values.  We use the latent factor structure of the high frequency data to fill in the `missing' values of the low frequency variable.  Since we  impute the weekly indicator using multiple factors, it is  conceptually distinct measure from the  WEI and the ADS index that are based on a single factor.    Treating high frequency observations as missing is not  new,\footnote{For example,  \citet{schorfheide-song:15, brave-butters-justiniano:19,itkonen-juvonen:17} consider BVARs. Weekly data are typically  seasonally unadjusted.} and neither is interpolation  between two data releases as we discuss below. However,  our $Y$  is itself a summary statistic  of economic activity such as the Chicago Fed National Activity Index (CFNAI), and constructing a  diffusion index  by solving  a missing data problem appears to be new. Modeling the problem in this way allows us to use  information in both frequencies  without  modeling a large number of variables at each frequency.  

Our aim is to  impute the missing high frequency observations. Imputation is concerned with filling in values that are never sampled and hence different from prediction and nowcasting which fills in values that are not yet observed but will eventually be observed. Many static matrix completion algorithms are available to impute missing  observations that are  assumed to be uniformly sampled, and static in the sense that   time dependence in the data can be ignored.  A  key assumption for successful recovery in such cases is that the complete data have a  'low-rank' component. In  the econometrics literature, the common factors and the loadings that determine the common component can be estimated from  incomplete data  in two ways.   The first  estimates the factors   jointly with the missing values. Methods in this class include  the EM algorithm in \citet{stock-watson-jasa:02} for estimation of static factors, and the  hybrid procedure   in \citet{dgr-joe} which also models the factor dynamics, which is  extended to allow for missing data in \citet{banbura-modugno:14}.   The second approach  separates estimation of factors from imputation of the missing values.  For example, the \textsc{tall-project (tp)} procedure in \citet{baing-miss:jasa} first estimates   the factors from a \textsc{tall} block of complete data  such as by  principal components (PC), quasi maximum likelihood (QMLE), or other consistent estimators.  Imputation is then based on the factor structure of the series of interest, treating the factor estimates as given.   Economic data are time dependent and it is known that the static factor estimates can be made more efficient if their  dynamics are modeled. However, while the properties of the factor estimates are well-studied, those of the imputed values, which are usually a by-product of factor estimation, are lesser known.


Given that our interest is  on the imputed values themselves, we start by exploring  to what extent the imputed values are affected by how the factors are estimated.
To this end, we first  evaluate several imputation methods on (quarterly, monthly) data when the missing monthly values occur in a systematic way.
Specifically,  we suppose that the consumer sentiment series  (CS) available only on a quarterly basis prior to 1978 was still observed quarterly after 1978 when monthly values were actually reported. We then consider  a counterfactual exercise that uses $N=122$ seasonally adjusted monthly  FRED-MD data to impute the artificially missing  monthly values.  We find  gains in estimation of the factor space by accounting for heteroskedasticity and the dynamics of the factors but only  when $N$ is very small, consistent with theory. Since $N$ in the consumer sentiment analysis is quite large,  existing methods  yield similar factor estimates, and consequently all impute similar missing values  that are unfortunately all bunched around the (normalized) unconditional mean of zero.
We suggest that  the  error in series to be imputed has information that can  improve the imputed values.


 Serial correlation in the idiosyncratic errors does not affect consistency of the factor estimates but has implications for prediction and imputation. Even though the errors have an unconditional mean of zero, accounting for  their persistence  can improve conditional prediction. We explore both single equation and systems  approaches to   {\it dynamic matrix completion}.    The first  single-equation estimator is that of \citet{chow-lin:71} which   combines  an infeasible Generalized Least Squares (GLS)  estimate of the mean with an estimate of the  error based on the persistent structure of the GLS residuals.  Further investigation indicates that this  best linear unbiased  estimator includes forward looking information in the prediction, akin to  the Kalman smoother.  This motivates a  \textsc{tp*} procedure that only uses the past  errors, akin to a filtered estimate, obtained by iteratively estimating  an unrestricted autoregressive distributed lag model with incompletely observed data. As for systems approach,  there are few methods that account for serial correlation in QMLE estimation with unstructured missing data  without enlarging the state space. We combine two state space models in the literature designed to solve two separate problems: serial correlation without missing data,  and structured missing values in the presence of serial correlation. 


Accounting for the residual correlation  delivers more precise imputed values not just in the counterfactual exercise on the consumer sentiment series, but also  in our  main application that imputes  weekly values  of the monthly index of economic activity, CFNAI.  Analysis of weekly data is more involved  because weekly seasonal variations are not strictly periodic. We assume that  the seasonal variations are  idiosyncratic and hence seasonally unadjusted data suffice for imputation. As in the analysis of the CS series,  the single equation procedures compete well with the more involved systems-based alternatives.  We identified episodes when the imputed weekly series convey information not in the monthly CFNAI that are consistent with economically-meaningful events that occurred.


The paper proceeds as follows.  Section 2 reviews  systems and single equation approaches to factor-based imputation.   Section 3 provides a counter-factual exercise to evaluate the imputation procedures when the data are missing at in a systematic way.  Methods that incorporate  dynamics in the idiosyncratic errors are discussed in Section 4. Section 5 turns to imputation of weekly data and the unique issues that arise. Our main message is  to pay attention to the treatment of idiosyncratic errors.  For some problems,  controlling  for serial correlation in the errors can be  far more important than the choice of the estimator for the factors. 



\section{Factor Based Imputation}

We are eventually interested in  filling in, or imputing,  the  values of a series $Y$  in the $m_t$ sub-periods (indexed by $s_t$)  between  $t$ and $t+1$ when  information in a large panel of data is also available. Research from various disciplines  recognized that a reduced rank  structure can facilitate recovery of the missing values.
While factor-based imputation exploits the common component,   {\it matrix completion} problems exploit the  low rank structure.  So whether one takes an algorithmic or statistical perspective,  estimation of the  common factors is needed.  

 We take the  factor model as a starting point. In the absence of missing values, the completely observed  data matrix is $X$, and  $X_{it}$, $i=1,\ldots N, t=1,\ldots T$  admit a factor representation
\[ X_{it}=\Lambda_i^\prime F_t+e_{it}.\]
A factor model decomposes the data into a common component $\Lambda_i^\prime F_t$ and an idiosyncratic component $e_{it}$, where $F_t$ is a 
vector of common factors of dimension $r<<N$, and $\Lambda_i$ is a $r\times 1$ vector of factor loadings. Classical factor models assume that $e_{it}$ is  mutually and serially uncorrelated so that  $\Sigma_e=\mathbb E[e_te_t']$ is an $N\times N$ diagonal matrix with $N$ fixed.  Following  \citet{baili:12,baili:16}, we let $\phi_t = \mathbb{E}\left(e_t e^\prime_t\right)$ while $\Phi = \text{diag}\left(\frac{1}{T}\sum_{t=1}^T \phi_t\right)$.


The issues in estimation of  classical factor models  from incomplete data  are similar to covariance structure modeling with missing data for which a larger literature is available. For a scalar series $y$,  let $y_{obs}$ denote the  observed values and $y_{miss}$ denote the  missing values. Corresponding to $y=(y_{obs},y_{mis})'$ of length $T$ is an indicator variable  $\Omega$, also of length $T$, whose value in the $t$-th position is  one if $y_{t}$ is observed and zero otherwise. Suppose that the  model parameterized by $\theta$ is $p(y|\theta)$ and the missing data mechanism is characterized by $p(\Omega|y;\phi)$. The  distribution of the observables is $p(y_{obs},\Omega|\theta,\phi)=\int p(y_{obs},y_{miss}|\theta)p(\Omega|y_{obs},y_{miss},\phi)dy_{miss}.$ There are two cases when  estimation of $\theta$ can proceed without modeling $\Omega$.  The  case of   {\it missing at random} (MAR) obtains when $p(\Omega|y_{obs},y_{miss},\phi)=p(\Omega|y_{obs},\phi)$ does not depend on $y_{miss}$. If, in addition, $p(\Omega|y_{obs},\phi)=p(\Omega|\phi)$ so that the density for $\Omega$ does not depend on $y_{obs}$,  then $y_{obs}$ is said to be observed at random (OAR) while $y_{miss}$  are   missing completely at random (MCAR)\footnote{For a formal definition, see \citet[Chapter 1]{little-rubin:19} and \citet{laird:88}.}. Since  missingness in the MCAR case is  independent of the data whether or not they are observed,   all values (observed or missing)   are  drawn from the same underlying  distribution. Hence rebalancing the panel by deleting cases  with missing data   will yield consistent estimates, albeit inefficient.

If the normality assumption  is correct, the maximum likelihood estimates of covariance structures have classical normal properties under MAR.   But as discussed in \citet{bentler:02},  the result holds only for MCAR when the data are  non-normal.  However,  \citet{arminger-sobel:90}  illustrate by means of an example that   assuming MCAR  could yield biased estimates when the data are actually MAR because  systematic differences between the observed and missing data are ignored.  The  missingness assumption is  not innocuous  even when  the  data  are independent and identically distributed, and  either $N$ or $T$ is fixed. This remains to be the case when $N$ and $T$ are large.

\subsection{Joint Estimation of the Factors and the Missing Values}


When $N$ and $T$ are both large, a simple estimator of the factors is  principal components (PC) whose  asymptotic properties  in    the complete data case are quite well understood. Assuming that the factor structure is strong, \citet{baing-ecta:02} show that  the factor space can be consistently estimated  while allowing the idiosyncratic errors to be weakly correlated.\footnote{Let $X=F\Lambda'+e$ be the factor representation of the $T\times N$ data matrix $X$. The factor structure  is said to be strong if $\Lambda'\Lambda/N$ and $F'F/T$ are both positive definite in the limit.} \citet{bai-ecta:03} establishes asymptotic normality of the estimated factors, the loadings, and the common component.  
The EM algorithm suggested in \citet{stock-watson-diforc} extends PC estimation to allow the data to be incompletely observed. Assuming MAR,  the algorithm first estimates  the factors   from an imputed data matrix by PC, and the estimated factors are then used to predict the missing values. The two steps are repeated until convergence. In spite of the popularity of the algorithm,   the statistical properties of the estimates are only  recently studied.    \citet{su-missing} formulate the problem by  reweighing the data  according to the missing propensity and show that  though the  estimates are consistent, they will  not be asymptotically normal without further iteration.   \citet{xiong-pelger:19} suggest an alternative approach that also  re-weighs the  data,   but iteratively estimates the static factors by  cross-section regressions. Their  `all purpose' estimator is robust to a variety of missing  mechanisms but is less efficient.
It is also possible to use mixed frequency data to improve the factor estimates as in the method of targeted principal components recently considered in  \citet{pelger-target:22}. These methods all impute missing values as a by-product, while  estimation of the factors is the goal.

State space models are set up to handle latent variables and can easily accommodate  incomplete data   if the errors are serially uncorrelated and missingness is random  so that   $\Omega$ does  not need to be  modeled.\footnote{With few exceptions such as in   \citet{durbin-koopman:02}   \citet{casella:13, cai-etal:19}, the  MAR assumption or its caveats are often not mentioned.} One typically starts with a  parameter driven model in which   the mean and variance are correctly specified and  that the innovations to the latent state are  unrelated to the observation error. This allows the missing values to  be replaced by their predicted ones implied by the Kalman filter while setting the Kalman gain of the missing observations  to zero. Filtering algorithms are  given in \citet{harvey-pierse-84}, \citet[Chapter 4]{durbin-koopman-book:12} and \citet[Chapter 6]{shumway-stoffer-book}, among others.  The idea is simple but can be  computationally costly as  missing values  increase  the dimension of the state spaces.\footnote{Efficient samplers are considered in \citet{chan-poon-zhu:21} and \cite{hauber-schumacher:21}.    See  \citet{bgmr-handbook:13} for a review.}  However, very few studies  have considered serially correlated noise in an incomplete data setting.

\subsection{Single Equation Imputation by \small{TP}}

\citet{baing-miss:jasa} suggest a new approach that  separates estimation of the factors from estimation of the missing values.  It does not assume MAR but instead  imposes additional assumptions on the factor model that characterizes the data.  It is best understood by looking  at the data matrix $X$ of dimension $T\times N$ rearranged into four blocks as shown below. Reorganizing the data is not necessary in practice but helps conceptualize the idea. The  \textsc{tall} block of dimension $T\times N_o$ consists of data for all $N_o$ units with data observed in every time period. The \textsc{wide} block of dimension $T_o\times N$ consists of data for each of the $N$ units  in a subsample of $T_o$ periods. The \textsc{bal} block is the intersection of \textsc{tall} and \textsc{wide}, and the remaining observations are collected into the \textsc{miss} block of dimension $(T-T_o)\times (N-N_o)$. As the size of this block is defined by $N_o$ and $T_o$,   there may be observed data inside  \textsc{miss}.

\begin{figure}[ht]
\caption{Reorganized data matrix of dimension $T\times N$:}

\label{fig:tallwide}

\begin{center}
\textsc{tall}=A+\textsc{bal}\\
 \textsc{wide}=B+\textsc{bal}

\includegraphics[width=6.0in,height=3.50in]{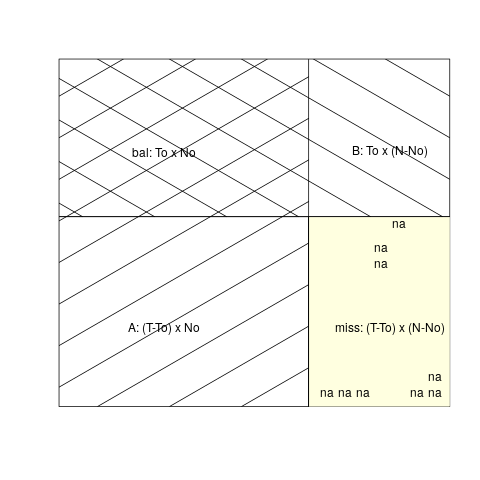}

\end{center}

\end{figure}

 The key insight   \citet{baing-miss:jasa} is that  $F$ and $\Lambda$  underlying the common component can be estimated from  two blocks of completely observed data without reference to the missing data mechanism. In particular,  the $T\times r$ matrix of  factors  can be consistently  estimated from  the \textsc{tall} block by PC.\footnote{ There are two different ways to estimate $\Lambda$.
 The \textsc{tw} algorithm in \citet{baing-miss:jasa} estimates $\Lambda$ from the \textsc{wide} block  and then aligns the estimate $\tilde \Lambda_{wide}$ with $\tilde F_{tall}$ through estimation of a rotation matrix. This method is better suited for missing data that occur in a structured way. The \textsc{tp} algorithm in \citet{bcn:21} estimates $H\Lambda_i$ directly using the subsample for which $i$ is observed.  This method  accommodates heterogeneous  missing patterns by having series specific sample size $T_{oi}$, and recognizing that the number of series observed each period ($N_{ot})$ can vary with $t$. As \textsc{tw} is a special case of \textsc{tp}, we will simply  refer to them as static matrix completion,  \textsc{tp}.}  Since this only involves complete data estimation,  PC  can be replaced by PC-GLS or other consistent estimators, including maximum-likelihood. As discussed in \citet{baing-miss:jasa}, consistency of $\tilde C_{it}$ for $C_{it}$ requires  that  (i) the factor structure is strong in each block and as a whole, (ii) that the blocks are sufficiently similar, and (iii) strict stationarity with the same population moments in all blocks.   Though the factor estimates are
 consistent without iteration,  one re-estimation of $F$ and $\Lambda$ by applying PC to  $\tilde X$  can accelerate the convergence rate of the estimated  $ C_{it}$ from $(\sqrt{N_o},\sqrt{T_{oi}})$ to $\sqrt{N_{ot}},\sqrt{T_{oi}})$ since $N_o\le N_{ot}$. An appeal of the approach is that since  $F$ and $\Lambda$ are  estimated from completely observed data, a  normal  distribution theory for the imputed values can be obtained. Treating potential outcomes as missing values and assuming that the errors are  serially uncorrelated, \citet{baing-miss:jasa}   obtain  inferential  results for the estimated  average treatment effect.

 We will apply \textsc{tp} to a mixed-frequency setting where of interest is   $Y$, a high frequency  series of length  $T=\sum_{t=1}^{T_o} m_t$,  $m_t$ being the number of sub-periods between $t$ and $t+1$.      The observed values of $Y$ correspond to those of the  low frequency  series  $y$ that is $T_o\times 1$. 
 Though $Y$  is incompletely observed, it admits  a factor representation
\begin{equation}
\label{eq:YH}
 Y= F\Lambda_Y+e_{Y}.
\end{equation}
Using $s_t=t+j/m_t$ for $j\in [1,m_t]$ to index the data in the higher frequency, we can also write
$  Y_{s_t}=\Lambda_Y' F_{s_t}+e_{Y,s_t}.$
Without loss of generality, we will assume that  data are released in the last sub-period between $t$ and $t+1$.

To  impute the missing values in $Y$, \textsc{tp}  exploits   a $T\times N_o$ matrix $X_o$   assumed to have  a factor structure $X_o=F\Lambda_{X_o}+e_{X_o}$. Hence in the \textsc{tp} setup,  $X=(X_o,Y)$.  Complete data  estimation of $F$ from $X_o$ (here, \textsc{tall}) yields  $\tilde F$.
The \textsc{wide} estimation  reduces to regressing the low frequency series $y$ on the subvector of $\tilde F$ consisting of the last sub-period at every $t$.  Assuming that    $e_{Y}$ is unpredictable,     \textsc{tp} will  return
\begin{equation}
\label{eq:tp-impute} 
\tilde Y_{s_t} = \begin{cases}  y_{t+1}  \quad\quad \quad\quad&   s_t=t+1 \\
\tilde \Lambda_{Y}' \tilde F_{s_t} & s_t=t+j/m_t, \quad j\in[1,m_t) \end{cases}. \end{equation}
According to \textsc{tp}, \textsc{tall} and \textsc{wide} suffice for imputation so long as $e_Y$ is serially uncorrelated.

\section{Imputing Consumer Sentiment using the Static Procedures }
We eventually want to  interpolate  weekly values of a monthly economic indicator. But as weekly data  are not regularly spaced and weekly panels are typically   smaller in size, we  first test  the procedures on regularly spaced   monthly data which are abundantly available.  Evaluating the adequacy of the interpolation methods  is challenging when the  values of interest  were never released.
 In this section, we take advantage of the change in the release of one series to perform such an evaluation.

   We consider  a hypothetical exercise using  the University of Michigan's index of consumer sentiment  (CS) as target.\footnote{This   series is available at \url{  https://fred.stlouisfed.org/series/UMCSENT.}}  This series  was  available on a quarterly basis from 1960 to 1978 (specifically, in February, May, August, and November), but   has been released on a    monthly basis since 1978. The counterfactual exercise  assumes  that the CS series remained available on a  quarter basis after 1978 and uses the different methods to impute the (2/3)$\times 504$ artificial missing monthly values.  We first demean and standardize the  CS series and then remove two observations each quarter in the early sample. The only series with missing values in the exercise is CS.  The complete data matrix  $X_o$ is  a balanced monthly panel of 122 series taken from  the December 2021 vintage of FRED-MD.

For  joint estimation of the factors and missing values, we implement a Kalman smoother (KS) and the EM procedure of \citet{stock-watson-diforc}. We also  consider several  estimators of $F$ for use in \textsc{tp}. To be clear about the exercise, different estimators of $F$ are used in \textsc{tp}, but the second (imputation) step  is the same.
These five estimators for $F$ are well known, we only summarize them in the Appendix. A detailed review is given in \citet{barigozzi-luciani:19}.

\begin{itemize}
\item[(i)] the PC  estimator  $(\tilde F,\tilde\Lambda)=(\sqrt{T} U_r, \sqrt{N}D_rV_r)$  where $U_r$ and $V_r$ are the left and right singular vectors corresponding to the $r$ largest singular values collected in the diagonal entries of $D_r$. The PC estimates minimize $\text{tr}[(X-F\Lambda')'(X-F\Lambda)]$.

\item[(ii)]   Assuming joint normality of $X$ and $F$,   
an  infeasible GLS estimator for $F$   which  accounts for heteroskedasticity
 is:
\begin{equation}
\label{eq:gls-F}  F^{GLS}_{s_t}=(\Lambda' \Phi^{-1}  \Lambda)^{-1}\Lambda' \Phi^{-1} X_{s_t}.
\end{equation}  
 The MLE-h replaces $\Lambda$ and $\Phi$ with the maximum likelihood estimates analyzed in \citet{baili:12}.  

\item[(iii)-(iv)]   \citet{breitung-tenhofen:11} use PC-GLS-har to account for both heteroskedasticity and serial correlation. We also consider a
 PC-GLS-h version that only controls for heteroskedasticity by updating the estimate of $F$  once using  the PC estimates of $\Lambda$ and $\Phi$.

\item[(v)] An infeasible projections   estimator that improves upon $\tilde F$  by using a VAR to  model its dynamics is
  $F^{p}_t= \mathbf U_t'\mathcal F^p$    where $\mathbf U_t'= (\iota_t\otimes I_r) $, $ \iota_t$  is the $t$-th column of the identity matrix $I_T$,
\begin{eqnarray} \mathcal F^{p}= E_\Theta(\mathcal F|\mathcal X]&=&\bigg(\Sigma_{\mathcal F}^{-1}+(I_T\otimes \Lambda'\Phi^{-1}\Lambda)\bigg)^{-1}\bigg( (I_T\otimes \Lambda' \Phi)\bigg) \mathcal X,
\label{eq:project-F-var}
\end{eqnarray}
$\mathcal F=\text{vec}(F')$, $\mathcal X=\text{vec}(X)$.
The feasible estimator (PC-KS) uses the Kalman smoother to construct the required terms.
We use the code implemented in \citet{barigozzi-luciani:19}.
\end{itemize}

\begin{center}
\begin{figure}[ht]
\caption{Monthly Consumer Confidence: Static} 

\label{fig:cs-static}

\includegraphics[width=6.50in,height=3.5in]{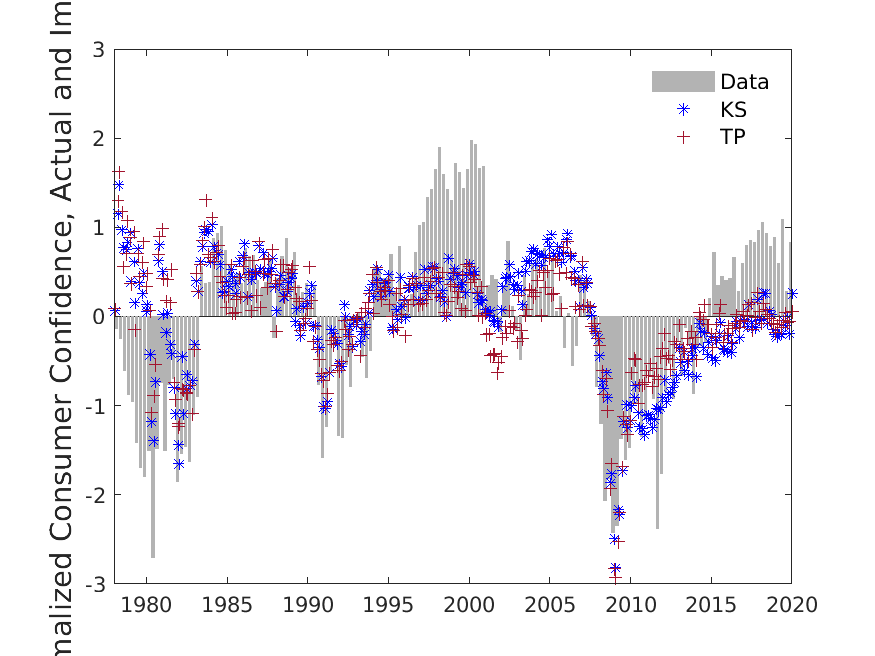}
\end{figure}
\end{center}

 Figure \ref{fig:cs-static} plots the normalized (demeaned and standardized) actual monthly consumer sentiment series (in grey) along with the series  imputed by \textsc{tp} (red plus sign), and \textsc{ks} (blue asterisk). The series using \textsc{mle-h}, \textsc{pc-gls-h}, \textsc{pc-gls-har}, \textsc{em}, and \textsc{pc-ks} are not shown because they are similar to \textsc{tp} and \textsc{ks}. By construction, the imputed values coincide with the observed values once a quarter (hence not displayed as they coincide with the true values in gray). But  notably,  the   imputed values   bunched around zero which suggest that the missing values are not well predicted by the factor model.  All  methods significantly underestimate variability of the series since the imputed series have a standard deviation of only 0.75, much below that of  the observed series, which is  one. Re-estimation of the factors from the complete data matrix  yields no noticeable improvements.  Since \textsc{tp} and \textsc{ks} give similar imputed values, the problem appears  not to be with joint or separate  estimation of the factors and missing values.   These results raise two questions. Why do different estimators of $F$ produce very similar results, and why are there episodes when   estimates of missing values are bunched around zero?

\subsection{A Calibrated Monte Carlo Exercise}

We consider two calibrated  experiments to understand why the choice of the estimator for the factors does not seem to matter. The first is calibrated to  a balanced panel of $T=720$ observations for $N$= 122 series over the sample 1960:1-2019:12 in the 2021-12 vintage of  FRED-MD, \citet{fred-md}.   The parameters used in the simulations are  based on PC estimation of $r=3$ factors. In this data, $\tilde F_1$  loads heavily on real activity variables and account for 0.147 of the total variation in the data, $\tilde F_2$   loads heavily on interest rates and account for  .073, while $\tilde F_3$ loads heavily on prices and account for  0.070 of the total variation in the data.  The implied idiosyncratic error variance $\tilde \phi_i$ range from  0.065 to  0.984, showing significant heteroskedasticity. The idiosyncratic errors are also serially correlated with $\rho_i$ ranging from -0.617 to 0.955. Estimation of a  VAR(2) in the three factors finds  the residuals to be mutually correlated. While $F_2$ has little serial correlation,  $F_1$ is quite persistent.

The  parameter estimates are used to simulate 1000 sets of data each with  Nsim series of length  $T=720$ by  taking at each $t$, (i) $N$  independent  normal draws of $e_{it}$, and (ii) $r$  normally distributed common shocks  while holding the loadings  fixed at the PC estimates.   To ensure that the simulated data are representative of the original  panel, the original panel is divided in three blocks of size (.4,.3,.3)$\times $ 122. Then  (.4,.3, .3)$\times$ Nsim series are simulated from each of the three blocks. We vary Nsim  from 10 to 122.

  \begin{table}[ht!]
  \label{tbl:table1a}
  \caption{DGP 1:  $ T=720, N=122, r=3$}

  \begin{center}
    \begin{tabular}{llllllllll}
  & \multicolumn{5}{c}{$M(F^*,\hat F)$}\\ 
       Nsim & 10 & 20 & 30 & 40 & 50 & 75 & 122 \\ \hline 
           PC &  0.669 &   0.703 &   0.868 &   0.902 &   0.875 &   0.923 &   0.957 \\
        MLE-h &  0.736 &   0.755 &   0.890 &   0.920 &   0.901 &   0.935 &   0.964 \\
     PC-GLS-h &  0.774 &   0.785 &   0.890 &   0.921 &   0.901 &   0.935 &   0.964 \\
   PC-GLS-har &  0.761 &   0.773 &   0.883 &   0.919 &   0.897 &   0.934 &   0.964 \\
           KS &  0.759 &   0.789 &   0.895 &   0.922 &   0.909 &   0.936 &   0.964 \\
 \hline 

    \end{tabular}
  \begin{center}
    Regression: $F^*_j=a+b_1 \hat F_1 + b_2 \hat F_2+ b_3\hat F_3$+ error
    \begin{tabular}{llllllllll}
      Nsim & 10 & 20 & 30 & 40 & 50 & 75 & 122 \\ \hline 
 $R^2: j=$1\\ \hline 
           PC &  0.736 &   0.841 &   0.897 &   0.936 &   0.922 &   0.956 &   0.971 \\
        MLE-h &  0.793 &   0.872 &   0.919 &   0.947 &   0.935 &   0.962 &   0.976 \\
     PC-GLS-h &  0.817 &   0.874 &   0.920 &   0.947 &   0.935 &   0.962 &   0.976 \\
   PC-GLS-har &  0.790 &   0.856 &   0.916 &   0.946 &   0.933 &   0.961 &   0.976 \\
           KS &  0.774 &   0.877 &   0.926 &   0.949 &   0.941 &   0.962 &   0.976 \\
 \hline 
 $R^2: j=$2\\ \hline 
           PC &  0.628 &   0.614 &   0.868 &   0.877 &   0.861 &   0.907 &   0.954 \\
        MLE-h &  0.713 &   0.686 &   0.886 &   0.903 &   0.896 &   0.924 &   0.962 \\
     PC-GLS-h &  0.769 &   0.741 &   0.885 &   0.905 &   0.897 &   0.924 &   0.962 \\
   PC-GLS-har &  0.769 &   0.740 &   0.879 &   0.901 &   0.893 &   0.923 &   0.962 \\
           KS &  0.770 &   0.743 &   0.888 &   0.905 &   0.902 &   0.924 &   0.962 \\
 \hline 
 $R^2: j=$3\\ \hline 
           PC &  0.642 &   0.655 &   0.837 &   0.891 &   0.842 &   0.905 &   0.946 \\
        MLE-h &  0.700 &   0.706 &   0.864 &   0.910 &   0.870 &   0.920 &   0.954 \\
     PC-GLS-h &  0.736 &   0.738 &   0.864 &   0.911 &   0.870 &   0.920 &   0.954 \\
   PC-GLS-har &  0.722 &   0.722 &   0.853 &   0.909 &   0.864 &   0.919 &   0.954 \\
           KS &  0.729 &   0.745 &   0.871 &   0.912 &   0.883 &   0.921 &   0.954 \\
 \hline 

    \end{tabular}
  \end{center}
  \end{center}
  {\footnotesize Notes: The top panel reports 
  \[M(F^*,\hat F)=\frac{ \text{trace}(F^{*'}\hat F(\hat F'\hat F)^{-1} \hat F'F^*)}{\text{trace}(F^{*'}F^*)},\]
  which  measures  the multivariate correlation between $F^*$ and their estimates $\hat F$. The bottom panel reports the $R^2_j$ from each $F_j^*$ regressed on $\hat F_1,\ldots, \hat F_r$. $F_j^*$ is estimated according to procedures listed in section 3: (i) PC; (ii) MLE-h; (iii) GLS-h; (iv) GLS-har; (v) KS.}
  \end{table}

Two sets of results are given in Table \ref{tbl:table1a}.  The top panel reports 
\[M(F^*,\hat F)=\frac{ \text{trace}(F^{*'}\hat F(\hat F'\hat F)^{-1} \hat F'F^*)}{\text{trace}(F^{*'}F^*)},\]
which  measures  the multivariate correlation between $F^*$ and their estimates $\hat F$. The PC estimates have noticeably larger errors  when Nsim is smaller than 50. However, even when Nsim is not small, there is little to gain in modeling dynamics of $F$ or of $e$ over controlling for heteroskedasticity.

Next,  each $F_j^*$ is regressed on $\hat F_1,\ldots, \hat F_r$ and the regression  $R^2_j$  is used as indicator of  how well information in $F^*_j$ is captured by $\tilde F$. These $R^2_j$s are reported in the bottom panel of Table \ref{tbl:table1a}. Three results are of 
note.  First, the improvements over PC are larger for estimates of $F_2$ and $F_3$ than for $F_1$.   Second, while the improvements in $R^2$ are noticeable when Nsim=10, they diminish rapidly   when Nsim exceeds 30. Third, and reinforcing the observation made above,  the gains of modeling the dynamics of $F$  over a GLS correction for heteroskedasticity are quite small. 

We also consider a second experiment that restricts the panel to  the first $N= 50$ variables in the panel which are all on real activity variables,   over a shorter sample that  starts in 1978 with $T=420$ observations. This time, the parameters used to simulate data are estimated by maximum likelihood. The first factor in this data is also  a real activity factor and all methods estimate it very well. As seen  in Table \ref{tbl:table1b}, the maximum likelihood estimates of $F_1$ are only slightly better than the PC estimates when Nsim=10. The efficiency gains in estimating $F_2$ are more noticeable. The gains from using the Kalman smoother are large at Nsim=10 but become much smaller when Nsim=30.

  \begin{table}[ht!]
  \caption{DGP2: $T=420, N=50$, $r=2$}
  \label{tbl:table1b}
  \begin{center}
  
  \begin{tabular}{llllllll}
  & \multicolumn{5}{c}{$M(F^*,\hat F)$}\\ 
    
 Nsim & 10 & 20 & 30 & 40 & 50 \\ \hline 
           PC &  0.602 &   0.879 &   0.913 &   0.936 &   0.946 \\
        MLE-h &  0.647 &   0.946 &   0.946 &   0.970 &   0.975 \\
     PC-GLS-h &  0.667 &   0.961 &   0.950 &   0.974 &   0.977 \\
   PC-GLS-har &  0.679 &   0.961 &   0.948 &   0.972 &   0.975 \\
           KS &  0.744 &   0.965 &   0.953 &   0.974 &   0.977 \\
 \hline 

  \end{tabular}
  \end{center}
  \begin{center}
  Regression: $F^*_j=a+b_1 \hat F_1 + b_2 \hat F_2+ b_3\hat F_3$+ error
  
  \begin{tabular}{lllllll}
     Nsim & 10 & 20 & 30 & 40 & 50 \\ \hline 
 $R^2: j=$1\\ \hline 
           PC &  0.820 &   0.926 &   0.947 &   0.962 &   0.970 \\
        MLE-h &  0.843 &   0.964 &   0.965 &   0.981 &   0.984 \\
     PC-GLS-h &  0.850 &   0.973 &   0.966 &   0.983 &   0.985 \\
   PC-GLS-har &  0.854 &   0.973 &   0.965 &   0.981 &   0.984 \\
           KS &  0.880 &   0.976 &   0.969 &   0.983 &   0.986 \\
 \hline 
 $R^2: j=$2\\ \hline 
           PC &  0.428 &   0.841 &   0.886 &   0.914 &   0.927 \\
        MLE-h &  0.491 &   0.932 &   0.930 &   0.962 &   0.967 \\
     PC-GLS-h &  0.522 &   0.952 &   0.936 &   0.966 &   0.970 \\
   PC-GLS-har &  0.538 &   0.952 &   0.934 &   0.964 &   0.968 \\
           KS &  0.637 &   0.956 &   0.940 &   0.967 &   0.971 \\
 \hline 

  \end{tabular}
  \end{center}
 {\footnotesize Notes: The top panel reports 
  \[M(F^*,\hat F)=\frac{ \text{trace}(F^{*'}\hat F(\hat F'\hat F)^{-1} \hat F'F^*)}{\text{trace}(F^{*'}F^*)},\]
  which  measures  the multivariate correlation between $F^*$ and their estimates $\hat F$. The bottom panel reports the $R^2_j$ from each $F_j^*$ regressed on $\hat F_1,\ldots, \hat F_r$. $F_j^*$ is estimated according to procedures listed in section 3: (i) PC; (ii) MLE-h; (iii) GLS-h; (iv) GLS-har; (v) MLE-ks.}
\end{table}

To understand why  the alternatives to PC perform so similarly, we need to understand the properties of projections estimators.   \citet{baili:12} consider a projections estimator that only controls for heteroskedasticity: 
\begin{equation}
\label{eq:project-F}  F_{t}^p=E_\Theta[F_t|X_1,\ldots,X_T]=\bigg(\Sigma^{-1}_{F}+ \Lambda'\Phi^{-1}  \Lambda\bigg)^{-1}  \Lambda' \Phi^{-1} X_{t}.
\end{equation}
 A Woodbury inversion\footnote{$(A+B)^{-1}=B^{-1}-(A+B)^{-1}AB^{-1}$ with  $A=\Sigma_{F}^{-1}$, $B=\Lambda'\Phi^{-1}\Lambda$.} of $\mathcal G= (\Sigma_{F}^{-1}+\Lambda'\Phi^{-1} \Lambda)^{-1}$  yields:
\begin{eqnarray*}
 F_{t}^p
&=&  (\Lambda'\Phi^{-1} \Lambda)^{-1} \Lambda'\Phi^{-1} X_t-\mathcal G
  \Sigma_{F}^{-1} 
(\Lambda'\Phi^{-1}\Lambda)^{-1}\Phi^{-1} X_{t}.
\end{eqnarray*}
But the first term  is simply the infeasible GLS estimator and  the second term converges to zero in mean square under the assumptions of a strong factor model. This leads to the result in  \citet{dgr-restat} and \citet{baili:16} that
  \[ F^p_{t}=F_{t}^{GLS}+O(N^{-1/2})\]
which is to say that the projections estimator is asymptotically equivalent to the GLS estimator that only controls for heteroskedasticity.

Now consider PC-KS  that also accounts for the dynamics of $F$.   A
     Woodbury inversion of $\mathcal G=\Sigma_{\mathcal F}^{-1}+(I_T\otimes \Lambda'\Phi^{-1}\Lambda)$ yields
\begin{eqnarray*}
F^{p}_{t}&=& F^{GLS}_{t}-\mathbf U_t'\mathcal G \Sigma^{-1}_{\mathcal F} \mathcal F
- \mathbf U_{t}'\mathcal G \Sigma_{F}^{-1} (I_T\otimes (\Lambda'\Phi^{-1} \Lambda)^{-1}) \Lambda'\Phi^{-1} \mathcal E
\end{eqnarray*}
where  $\mathcal E=\vec(e)$ is $NT\times 1$.
Consistency of the (infeasible) projections estimator for $F_{t}$ follows because $F_{t}^{GLS}\pconv F_{t}$ and the second term vanishes at rate $\sqrt{N}$. The feasible PC-KS estimator is also consistent for $F$. While consistency of the estimator is recognized, less appreciated is  that   it is  asymptotically equivalent to  the GLS estimator that only controls for heteroskedasticity. Indeed,  \citet{baili:16}  show that  $\hat F^{PC-KS}_{t}-\hat F^{GLS}_{t}=O_p(N^{-1/2})+O_p(T^{1/2}N^{-2})$ if $T/N^3\rightarrow 0$, where $\hat F_{t}^{KS}$ and $\hat F_{t}^{GLS}$ are the QMLE estimates of  $\Lambda$, $\Phi$, $\Sigma_{\mathcal F}$.

An overlooked feature of   projections estimators is that they are  shrinkage estimators that shrink towards $F_{s_t}^{GLS}$. The degree of shrinkage  depends  on the relative importance of each factor and will be larger for  factors with  smaller variances.  As seen from Table \ref{tbl:table1a},  the less important factors stand to gain more from shrinkage.  Furthermore,  to the extent that $F_{s_t}^{GLS}$ is unbiased, the shrinkage estimators  will be biased, but the  smaller variance may result in  a more favorable mean-squared error.   Under the strong factor assumption, $\Lambda'\Phi^{-1}\Lambda$ is $O_p(N)$ while  $\Sigma^{-1}_{F}$ does not increase with $N$.\footnote{This contrasts with the rank regularized (RPC) estimator in \citet{baing-joe:19} that uses iterative ridge regressions to threshold singular values less than a predetermined value, but  does not weigh $\Lambda'\Lambda$ by $\Phi^{-1}$.}  Hence, the shrinkage factor  will vanish as $N$ increases. Our calibrated Monte Carlo simulation results are consistent with theory. Controlling for  heteroskedasticity in the idiosyncratic errors generates the largest  improvements over the PC estimates of $F$.

In summary, whether the idiosyncratic errors are serially correlated or heteroskedastic,  estimators more efficient than PC are available.   The above Monte Carlo  experiments confirm that  gains do  appear in finite samples. But   in the empirical exercise when $N$ is quite large, all consistent estimators of  $F$ produce similar imputed values. Though these will be unbiased provided  $\mathbb E[e_t]=0$, they  are unfortunately  bunched around the normalized unconditional mean of zero. But as \citet{phillips-joe:79} noted, 
prediction is made  given the final value of the endogenous variable in practice. In the next section, we will argue that the bunching arises when  the realized $e_{T}$  is not  zero. We then suggest different ways to incorporate the information into the imputation.  

\section{Dynamic Imputation}

This section considers three ways to account for residual serial correlation in imputation given related covariates. The first is the seminal work of \citet{chow-lin:71}. The second is an iterative algorithm that  estimates  a factor-augmented autoregressive distributed lag model from incomplete data. The third  is joint modelling of the latent factors and missing values while accounting for residual serial correlation in a fully specified state space system. The first and third methods produce smoothed estimates that  use  information beyond the period that the missing value occurs. The third method is fully specified and can be more efficient but  is computationally costly. Thus, each method has some appeal.

\subsection{The Chow-Lin Procedure}
\citet{chow-lin:71} provide a methodology for best linear unbiased  extrapolation and interpolation of a time series using a small set of related variables.  The analysis is built on the classic result of \citet{goldberger-62}  that analyzes the case   when complete data are available for estimation of a linear prediction model $Y=Z\beta+u$ in which  the errors are  non-spherical.  The best linear unbiased  prediction (BLUP) of some $Y_{T+\ell}$ when predictors $Z_{T+\ell}$ are available takes the form    $\hat  Y_{T+\ell}= Z_{T+\ell}'\hat \beta+ \mathbb E[u_{T+\ell} u']' V^{-1} \hat u$, where $\hat u=Y-Z\hat\beta$ is the vector of in-sample residuals, and $\hat\beta$ is the GLS estimator.
In the AR(1) case when  $u_t=\rho u_{t-1}+\epsilon_t$, the BLUP simplifies to
$ \hat Y_{T+\ell }=  Z_T'\hat\beta+\rho^{\ell } u_T$.   The equivalent representation $\hat Y_{T+1}=Z_{T}'\hat\beta+\rho(y_T-Z_T\hat\beta)$ for $\ell=1$  was suggested in  \citet{cochrane-orcutt:49} as a  way to improve prediction efficiency when the errors are correlated.  Note that though the omitted term $\rho^{\ell} u_T$ does not create unconditional bias so long as $\mathbb E[u_T]=0$, omitting the second term can induce bias  if the $u_T$ in  the  data is non-zero. 


The Chow-Lin analysis is more complicated because  data  of different frequencies are involved  and a matrix $C$ is needed to map the data from one frequency to another. The target is  a vector of non-sampled values, say, $q_*$.  For interpolating stock variables,  each row of  $C$  would have  a single  non-zero entry to indicate which subperiod  is observed, while  for  flow data, each row of $C$  would have $m_t$ non-zero entries to  pick the sub-periods that are to be aggregated. The \citet{goldberger-62} analysis is a special case when  $C$ is an identity matrix. To make our point clear, we only consider stock variables which is also notationally simpler.

We will subscript variables by $H $ or $L$  when we are referring to the Chow-Lin procedure.  Let  $Z_H$ be a $T_H\times K$ matrix of observed high frequency predictors\footnote{For our application we replace $Z_H$ with $\hat{F}_s$, but for this description we keep the original notation of $Z_H$. Chow and Lin assumed $Z_H$ observed.} for $Y_H$ modeled as
\begin{equation*}
 Y_H=Z_HB+u_H, \quad\quad u^H\sim (0,V_H).
\end{equation*}
Multiplying both sides by the $T_L\times T_H$  predetermined matrix $C$ gives the {\it bridge equation}
\begin{equation}
\label{eq:bridge}
 Y_L=Z_L B+ u_L,\quad\quad u_L\sim (0, V_L)
\end{equation}
where   $V_L=CV_HC'$, $C Y_H=Y_L$,   $C Z_H=Z_L$ and $u_L=Cu_H$.  The goal is to obtain linear unbiased imputation of  $q_*$ when $u_*=u_H$. To do so, \citet{chow-lin:71} consider linear predictors of the form  by  $\hat q_*=AY_L=A(Z_L+u_L)$  where $A$ is    $T_*\times T_L$, and  minimize trace(\text{cov}$(\hat q_*q_*))$ with respect to $A$ subject to the linear-unbiased constraint that $AZ_L-Z_*=0$.   Though $B$  is identifiable from the low frequency bridge equation,   
terms do not  simplify as in \citet{goldberger-62} because $C$ is  no longer  an identity matrix.   Nonetheless, the  (infeasible) BLUP of  $q_*$  has a familiar form:
\begin{equation}
\label{eq:chowlin}
 \mathbb E[ q_*|Z_H,Y_L]= Z_{H*}' \hat\beta_{GLS}+ \mathbb E[u_* u_L']' V_L^{-1}\hat u_L
\end{equation}
where $\hat\beta_{GLS}$  estimates $\beta= (Z_L' V_L^{-1} Z_L)^{-1} Z_L'V_L^{-1} Y_L$. 
As in \citet{goldberger-62},  the solution has two components. The first combines the high frequency information $Z_H$ as predictors with GLS estimates of  $\beta$ as weights to form the conditional mean. When $u_H$ is serially correlated, the second distributes the low frequency error $\hat u_L$ to the sub-periods between $t$ and $t+1$   under an assumed error structure. 
 \citet{chow-lin:71} assume an AR(1) model for $u_H$ while \citet{fernandez:81,litterman:83} considers  higher order persistence. Though the $V_H$ matrices  differ,  the idea is the same.


In spite of the popularity of the Chow-Lin procedure,  there is little discussion of  how $\hat u_L$ is actually distributed in the remaining $m_t-1$ sub-periods. Note that $q^*$ is not sampled and thus unobserved after the fact, which complicates evaluation of procedures.  With the help of some symbolic math calculations, we find that  for AR(1) errors parameterized by $\rho$, the procedure produces
\begin{eqnarray*}
\hat Y_{H,s_t}&=&\hat\beta_{GLS}'Z_{H,s_t} + \hat\theta_{1,s_t} \hat u^L_t+\hat \theta_{2,s_{t+1}} \hat u^L_{t+1}\\
&=&\hat\beta_{GLS}'Z_{H,s_t} + \hat\theta_{1,s_t} \bigg(Y_{L,t}-\hat\beta_{GLS}'Z_{L,t}\bigg)+\hat \theta_{2,s_{t+1}} \bigg( Y_{L,t+1}-\hat\beta_{GLS}'Z_{L,t+1}\bigg)
\end{eqnarray*}
where the weights $\hat\theta_{1,s_t}$ and $\hat \theta_{2,s_{t+1}}$ depend on $\rho$ and are different at the endpoints from the interior of the sample.\footnote{If we observe every other data point, $Y_{H,t+1/2}=\theta_{1,s_t}=\frac{\rho}{\rho^2+1}(\hat u_{L,t}+\hat u_{L,t+1})$. If $m=3$, $\hat Y_{H,t+1/3}=\frac{\alpha^3+\alpha}{\alpha^4+\alpha^2+1}\hat u_{L,t}+\frac{\alpha^2}{\alpha^4+\alpha^2+1} \hat u_{L,t+1}$ and $Y_{H,t+2/3}=\frac{\alpha^2}{\alpha^4+\alpha^2+\alpha} \hat u_{L,t}+ \frac{\alpha^3+\alpha}{\alpha^4+\alpha^2+1}\hat u_{L,t+1}$.} The interesting feature is that  in the $m_t-1$  sub-periods when $Y_{H,s_t}$ is missing,
 $\tilde Y_{H,s_t}$ will only depend on $\hat u_{L,t}$ and $\hat u_{L,t+1}$ but not values of $\hat u_L$ before $t$ or  after $t+1$. This is intuitive in retrospect because $s_t\in (t,t+1]$. By  using  information in $t+1$,  the Chow-Lin procedure obtains a  smoothed estimate of the serially correlated noise.  The non-zero weights also imply that the imputed series will be  smoother than when $\rho$ is assumed to be zero.

There are variations to the Chow-Lin procedure, many motivated by the  unsynchronized  release of  macroeconomic data towards the end of the sample. The so-called {\it ragged-edge} problem noted  in \citet{wallis:86}  has generated much interests in  extrapolating the target variable using data that are incompletely released, also known as nowcasting.  The MIDAS (mixed-data sampling) approach of   \citet{ghysels:06} replaces $Z_H B$ by a distributed lag of  $ Z_H$ with parameters $\delta_H$  to be  jointly estimated   with $B$ by constrained non-linear least squares, but without imposing the unbiased constraint.     \citet{schumacher:16} suggests that the difference between MIDAS and Chow-Lin type procedures is akin  to direct versus iterative forecasts.  The UMIDAS  of \citet{foroni-marcellino} estimates  $\delta_H$  freely. \textsc{tp} is most closely related to  UMIDAS with $Z_H$ replaced by $\tilde F$.

\subsection{\small{TP*}}

The \citet{chow-lin:71} procedure models  persistence in the residuals. 
\citet{santos-cardoso} rewrite  the model in terms of  an infinite order distributed lag in $Z_H$ that is eventually   approximated  by a finite number of lags.   \citet{ghysels-santaclara-valkanov:05} caution in the context of  MIDAS that this way of introducing  autoregressive dynamics in mixed frequency estimation  can generate  spurious periodic responses.  To circumvent this problem, \citet{clements-galvao:08} add  common autoregressive dynamics to both $Y_L$ and the predictors in  the  bridge equation. In the spirit of \citet{chow-lin:71}, lead and lagged information pertain to those of $Y_L$ through the bridge equation.

Why not directly use lags of $Y_H$? The challenge  is that $Y_H$ is incompletely observed. In spite of early   contributions from  \citet{sargan-drettakis:74,dunsmuir-robinson:81,palm-nijman:84}, the   literature on  estimating  dynamic models with missing data remains quite small. \citet{palm-nijman:84} consider identification issues in estimation of  ARMA models with missing data.  
 In the missing data literature developed mostly for i.i.d. data with focus on parameter estimation and not the missing values themselves, the generic procedure  is to use  {\it sweep} operations to first predict the missing values one variable at a time, irrespective of whether the variable  is a covariate or dependent variable. The output is then used to adjust the sufficient statistics for bias due to  missing values. As illustrated in  \citet[Ch. 11]{little-rubin:19}   for the AR(1) model where $y_1,y_3,\ldots, y_{T-1}$ are observed but not $y_2,y_4,\ldots, y_T$,     the adjustments require  an implicit regression of $y_t$ on $y_{t-1}, y_{t+1}$. Hence like the Chow-Lin procedure, forward information up to $t+1$ is incorporated without directly using the Kalman filter. However,  the  bias adjustments are model specific and the sweep operations can be cumbersome to implement.

We consider  an approach that is easy to implement but is less efficient because it uses only lagged  information. Suppose that   $u_{H,s_t}=\rho u_{H,s_t-1/m_t}+\epsilon^0_{H,s_t}$. Quasi differencing at $\rho$ gives
\[ Y_{H,s_t}=Z_{H,s_t}'\beta+\rho (Y_{H,s_t-1/m_t}-\beta' Z_{H,s_t-1/m_t})+\epsilon^0_{H,s_t}.\]
 When $\rho\ne 0$,  the lagged error of the static model, though mean zero, contains information to improve the prediction. 
The  unrestricted form of the above equation is the  equation in \citet{durbin:70}
\[ Y_{H,s_t}=\rho Y_{H,s_t-1/m_t}+\beta'Z_{H,s_t}+ \gamma'  Z_{H,s_t-1/m_t}+\epsilon_{s_t}.\]
The problem is  transformed from a bridge regression in completely observed low frequency data with a serially correlated error, to a regression in high frequency data with a serially uncorrelated error, but that   the dependent and lagged dependent variable  are incompletely observed.   

A naive approach is to use deterministic rules to interpolate the  missing values between $t$ and $t+1$, such as  replacing all missing  $y_{s_t}$  for $s_t\in (t,t+1]$ with the most recently observed value, $y_t$.  This can conveniently accommodate irregular missing patterns but   would not incorporate the available high frequency information between $t$ and $t+1$. We suggest to improve upon the naive approach by  iteratively updating the imputed series.  Starting at $k=0$, we  estimate  $\rho^{(k)}, \beta^{(k)} $ and $\gamma^{(k)}$ from
\begin{eqnarray*}
 Y^{(k)}_{H,s_t}=\rho Y^{(k)}_{H,s_t-1/m_t}+\beta' Z_{H}+ \gamma' Z_{H,s_t-1/m_t}+\epsilon_{s_t}.
\end{eqnarray*}
Then for those $s_t$ for which $Y_{H,s_t}$ are not observed,  we set  $Y^{(k+1)}$ to $\hat Y_{H,s_t}=\hat \rho^{(k)} Y^{(k)}_{H,s-1}+Z_{H,s-1}'\hat \beta^{(k)}+Z_{H,s-1}'\hat\gamma^{(k)}$ and update $k$ to  $k+1$.  Iteration stops when $|Y^{(k+1)}-Y^k|$ is less than some tolerance.   Iterative estimation of missing values  seems to date back to \citet{healy-westmacott:56} but is mainly used in i.i.d. settings.    In the two pen-and-pencil cases considered, we can show that  the converged  estimates are the true values and that a sign restriction on $\rho$  is needed  to ensure a unique solution, consistent with the result in  \citet{palm-nijman:84}. 

To help understand the difference between Chow-Lin and the iterative procedure, consider a simple example with $m_t=m=3$.  The iterative procedure  would  impute  $Y_{H,1+2/3}$  as
\begin{eqnarray*}
\hat Y_{H,t+2/3}&=& \hat \alpha \hat Y_{H,t+1/3}+ \hat\gamma_1'Z_{H,t+2/3}+\hat  \gamma_2' Z_{H,t+1/3}\\
&=& \hat\alpha\bigg( \hat \alpha Y_{H,t}+\hat\gamma_1'Z_{H,t+1/3}+\hat\gamma_2' Z_{H,t}\bigg)+\hat\gamma_1'Z_{H,t+2/3}+\hat\gamma_2' Z_{H,t+1/3}\\
&=& \hat\delta_Y' Y_{H,t}+ \hat\delta_{Z,0} Z_{H,t}+  \hat\delta_{Z,1}' Z_{H,t+1/3}+\hat\delta_{Z,2}'Z_{H,t+2/3},
\end{eqnarray*}
which is a weighted sum of  $Y_{H,t}$,  $Z_{H,t},  Z_{H,t+2/3}, Z_{H,1+t/3}$, and the $\delta$ parameters are estimated by OLS.   Using the fact that $Y_{L,t}=Y_{H,t}$,  $Z_{L,t}=Z_{H,t}$ 
 and $Z_{L,t+1}=Z_{H,t+1}$, the Chow-Lin procedure returns
\[ \hat Y_{H,t+2/3}=\hat\beta_{GLS}' Z_{H,t+2/3}+ \hat\theta_{1,t}(Y_{H,t}-\hat\beta_{GLS}' Z_{H,t})+ \hat \theta_{2,t}(Y_{L,t+1}-\hat\beta_{GLS}' Z_{L,t+1}),\]
which is a weighted sum of $Y_{H,t}, Z_{H,t}, Z_{H,t+2/3}$ {\it and} $  Y_{H,t+1}, Z_{H,t+1}$, and the parameters are estimated by GLS.  Notably, the iterative procedure does not include $t+1$ information, but it incorporates  $Z_{H,t+1/3}$ not used in Chow-Lin.

With this background, we can now modify \textsc{tp} to \textsc{tp*}
so that  \textsc{wide} step  is  iterative estimation of  an autoregressive distributed lag regression. Starting  at $k=0$ with  $Y_{s_t}^{(0)}=\tilde Y_{s_t}$ ,  estimate
\begin{eqnarray}
Y^{(k)}_{s_t}= \beta_0+ \rho Y^{(k)}_{s_t-1/m_t}+\Gamma_Y(L)'  \tilde F_{s_t-1/m_t} +w_{Y,s_t}.
\label{eq:fadl}
\end{eqnarray}
 The fit is then used to  update    $Y^{(k+1)}_{s_t}$   if $Y_{s_t}$  is missing at $s_t$. Separating the task of  estimating $F$ from  imputation provides a   way to control for serial correlation in $e_{Y,s_t}$ without changing estimation of $F$.       Note that while static \textsc{tp} does not require iteration,   \textsc{tp*}  does  require iteration to ensure that the $Y$ and lagged values used in the regression are internally consistent.   
It can be  used with any estimator of $F$ in the \textsc{tall} step.  

\subsection{A Modified Kalman Filter}

\def\kfL{\Lambda}
\def\kfF{\mathcal{F}}
\def\kfVAR{\mathcal{A}}
\def\epsS{\epsilon}
\def\epsSa{\tilde{\epsilon}}
\def\epsSi{\xi}
\def\etaS{\eta^*}
\def\t{s_t}

\textsc{tp*} is what \citet{bgmr-handbook:13} refer to as partial approach that do not specify a joint model for the variable of interest. A systems approach is more efficient but can be computationally costly. With this approach,
we model the data  $X_{s_t}=(X_{o,s_t}',Y_{s_t})'$  as being driven by latent factors $\kfF_{s_t}$ and serially correlated idiosyncratic errors $\epsSa_{i\t+1/mt}$ 
\begin{align*}  
  \text{Observation Equation:} && X_{\t} &= \kfL\kfF_{\t} + \epsS_{\t} && \\
  && \epsS_{i\t} &= \epsSa_{i\t} + \epsSi_{i\t} && \epsSi_{i\t} \sim N \left(0,\sigma^2_{\epsSi i} \right) \\
  &&\epsSa_{i\t+1/m_t} &= \rho_i\epsSa_{i\t} + e_{i\t+1/m_t} && e_{i\t} \sim N \left(0,\sigma^2_{ei} \right) \\ 
  \text{Transition Equation: } && \kfF_{\t+1/m_t} &= \kfVAR\kfF_{\t} + \eta_{\t} && \eta_{\t} \sim N \left(0,\Sigma_\eta \right)
\end{align*}
where  $\epsS_{\t} = \left(\epsS_{1\t}, \epsS_{2\t}, \epsS_{N_0\t} \epsS_{Y\t}\right)^\prime$,  $e_{s_t}$, $\xi_{s_t}$, and $\epsSa_{s_t}$ are similiarly defined.
The issue is that the standard Kalman filter and smoothing equations such as in \citet{durbin-koopman:02} cannot be applied because  some equations are not standard observation and transition form.  \citet{junbacker-koopman-vanderwei:11} provides a method to account for residual serial correlation in a state-space setup but requires that  we observe two sequential values of the target variable. The weekly data that we will subsequently consider are irregularly spaced.

In order to model the residual serial correlation in the presence of missing errors in a tractable manner, we combine two existing techniques of compressing the extra processes: quasi-differencing (Method A), and expanding the state space (Method B). Quasi differencing is computationally tractable for large datasets, but is not appropriate for series with missing data. Expanding the state space is theoretically feasible for datasets with missing data, but empirically impractical because of the computation cost - the number of latent states increases with the number of variables in the dataset. 
Our proposed approach is to quasi difference series for which data are completely observed, and only  add a series-specific predictable component to the state vector for our target variable which has missing values in the high frequency. 
We briefly review the two methods that are combined before presenting the proposed state space setup.  There are $r$ factors in the general case, $\kfF_t = \left(\kfF_{1t}, \kfF_{2t}, \dots, \kfF_{rt}\right)^\prime$. We use a two factor case as an example for Method B and our proposed procedure.

\paragraph{Method A: Accounting for Serial Correlation with Quasi Differencing}
This method follows \citet{watson-engle-83,reis-watson:10}. Define $\rho = \text{diag}\left(\rho_1, \dots, \rho_{N_0}, \rho_Y \right)$.  Quasi differencing  both the left and right hand side of the observation equation $X_{s_t}=\Lambda \kfF_{s_t}+\epsS_{\t}$ gives
\begin{align*}
  \left(\mathbb{I} -\rho L \right)X_{\t} &= \left(\mathbb{I} -\rho L \right)\left(\kfL\kfF_{\t} + \epsS_{\t} \right)&& \\
  X_{\t} - \rho X_{\t-1/m_t} &= \kfL\kfF_{\t} - \rho \kfL\kfF_{\t-1/m_t} + \epsS_{\t} - \rho\epsS_{\t-1/m_t}
\end{align*}
Note that $\epsS_{\t} - \rho\epsS_{\t-1/m_t} = e_{\t}$. Rearranging terms yields
\begin{align*}
  X_{\t} &= \underset{c_{\t}}{\underbrace{\rho X_{\t-1/m_t}}} + \underset{\kfL^A}{\underbrace{\left(\kfL, -\rho\kfL \right)}} \underset{\kfF_{\t}^A}{\underbrace{\begin{pmatrix} \kfF_{\t} \\\kfF_{\t-1/m_t} \end{pmatrix}}} + e_{\t}
\end{align*}
In state space formulation, the observation equation can be replaced with
\begin{align*}
  X_{\t} &= c_{\t} + \kfL^A\kfF_{\t}^A + e_{\t}
\end{align*}
Where $e_{\t}\sim N\left(0,\text{diag}\left(\sigma^2_{e1},\dots\sigma^2_{eN_0},\sigma^2_{eY}\right) \right)$, thus pre-removing serial correlation. The term $c_{\t}$ is known at time $\t$ and can be easily accounted for in the updating equations.

\paragraph{Method B: Accounting for Serial Correlation by Expanding the State Space}
This method follows \citet{banbura-modugno:14}. Consider a two factor model, $\kfF_{\t} = \left(\kfF_{1\t}, \kfF_{2\t}\right)^\prime$. The transition equation for the factors can be written as
\begin{align*}
  \begin{pmatrix}
    \kfF_{1\t+1/m_t} \\
    \kfF_{2\t+1/m_t}
  \end{pmatrix} = \begin{pmatrix}
    \kfVAR_{11} & \kfVAR_{12} \\
    \kfVAR_{13} & \kfVAR_{22}
  \end{pmatrix} \begin{pmatrix}
    \kfF_{1\t} \\
    \kfF_{2\t}
  \end{pmatrix} + \begin{pmatrix}
    \eta_{1\t+1/m_t} \\
    \eta_{2\t+1/m_t}
  \end{pmatrix}
\end{align*}

The autocorrelated part of the idiosyncratic errors $\epsSa_{i\t}$ can be added to the state space for all $N$ variables
\begin{align*}
  \underset{\kfF_{\t+1/m_t}^{B}}{\underbrace{\begin{pmatrix}
    \kfF_{1\t+1/m_t} \\
    \kfF_{2\t+1/m_t} \\
    \epsSa_{1\t+1/m_t} \\
    \epsSa_{2\t+1/m_t} \\
    \vdots \\
    \epsSa_{Y\t+1/m_t}
  \end{pmatrix}}}
 = \underset{\kfVAR^{B}}{\underbrace{\begin{pmatrix}
  \kfVAR_{11} & \kfVAR_{12} & 0 & 0  & \dots & 0 \\
  \kfVAR_{21} & \kfVAR_{22} & 0 & 0  & \dots & 0 \\
  0 & 0 & \rho_1 & 0  & \dots & 0 \\
  0 & 0 & 0 & \rho_2  & \dots & 0 \\
  0 & 0 & 0 & 0 & \ddots & 0 \\
  0 & 0 & 0 & 0 &  \dots & \rho_Y
\end{pmatrix}}} \underset{\kfF_{\t}^{B}}{\underbrace{\begin{pmatrix}
  \kfF_{1\t} \\
  \kfF_{2\t} \\
  \epsSa_{1\t} \\
  \epsSa_{2\t} \\
  \vdots \\
  \epsSa_{Y\t}
\end{pmatrix}}} + \underset{\eta_{\t}^{B}}{\underbrace{\begin{pmatrix}
  \eta_{1\t+1/m_t} \\
  \eta_{2\t+1/m_t} \\
  e_{1\t+1/m_t} \\
  e_{2\t+1/m_t} \\
  \vdots \\
  e_{Y\t+1/m_t}
\end{pmatrix}}}
\end{align*}
Define:
\begin{align*}
  \kfL^{B} \equiv \begin{pmatrix}
    \Lambda & \mathbf{0}_{N\times N} \\
    \mathbf{0}_{N\times N} & \mathbb{I}_N
  \end{pmatrix}
\end{align*}
Thus the observation equation can be rewritten in a form of a serially uncorrelated  idiosyncratic error
\begin{align*}
  X_{\t} &= \kfL^{B}\kfF^{B}_{\t} + \xi_{\t}
\end{align*}
where $\xi_{\t} \sim N \left(0,\text{diag}\left(\sigma^2_{\xi 1},\dots,\sigma^2_{\xi N_0},\sigma^2_{\xi Y} \right)\right)$.

\paragraph{Proposed KS*}
While quasi differencing the data is preferred as it does not increase the estimated state space, it can  be applied  only if we  observe two sequential values of the target variable. As  missing values can be irregularly spaced as in the case of weekly data, we write the state space system using a combination of the two methodologies described above: completely observed variables are quasi differenced, and the autocorrelation of the serially correlated error of the target series is included as an additional state. Consider again a two factor model. We can rewrite the transition equation as follows

\begin{align*}
  \begin{pmatrix}
    \kfF_{1\t+1/m_t} \\
    \kfF_{2\t+1/m_t} \\
    \epsSa_{Y\t+1/m_t}
  \end{pmatrix}
 = \underset{\kfVAR^{*}}{\underbrace{\begin{pmatrix}
  \kfVAR_{11} & \kfVAR_{12} & 0 \\
  \kfVAR_{21} & \kfVAR_{22} &  0 \\
  0 & 0 & \rho_Y
\end{pmatrix}}}\begin{pmatrix}
  \kfF_{1\t} \\
  \kfF_{2\t} \\
  \epsSa_{Y\t}
\end{pmatrix} + \underset{\eta_{\t}^{*}}{\underbrace{\begin{pmatrix}
  \eta_{1\t+1/m_t} \\
  \eta_{2\t+1/m_t} \\
    e_{Y\t+1/m_t}
\end{pmatrix}}}
\end{align*}
We can then take into account serial correlation of the completely observed series using quasi differencing in the observation equation. In this framework, however, define $\rho^* = \text{diag}\left(\rho_1, \dots, \rho_{N_0}, 0 \right)$. The observation equation is then

\begin{align*}
  X_{\t} &= \underset{c_{\t}}{\underbrace{\rho^* X_{\t-1/m_t}}} + \underset{\kfL^{*}}{\underbrace{\left(\kfL, -\rho^*\kfL, \begin{pmatrix} \mathbf{0}_{N_0\times 1} \\ 1 \end{pmatrix} \right)}} \begin{pmatrix} \kfF_{\t} \\\kfF_{\t-1/m_t} \\ \epsSa_{Y\t} \end{pmatrix} + \begin{pmatrix}
    e_{1\t} \\
    \vdots \\
    e_{N_0\t} \\
    \xi_{Y\t}
  \end{pmatrix}
\end{align*}

The state space comprises of the original latent factors, the lags of these factors, and one additional state for the predictable residual component. This combined approach remains computationally tractable for large $N$. In implementation, we modified the code from \citet{barigozzi-luciani:19}.\footnote{Estimates of $\rho_N$ are backed out from estimates of $\rho_N^{\frac{1}{\mathbf{s_o}}}$ where $\mathbf{s_o}$ is the share of values observed in the target series. For weekly data this is between 1/4th and 1/5th.}.

\subsection{Re-examining  the \small{CS}  Series}

\begin{center}
\begin{figure}[ht]
\caption{Monthly Consumer Confidence: Dynamic vs Static} 
\label{fig:cs-dynamic}
\includegraphics[width=6.50in, height=3.5in]{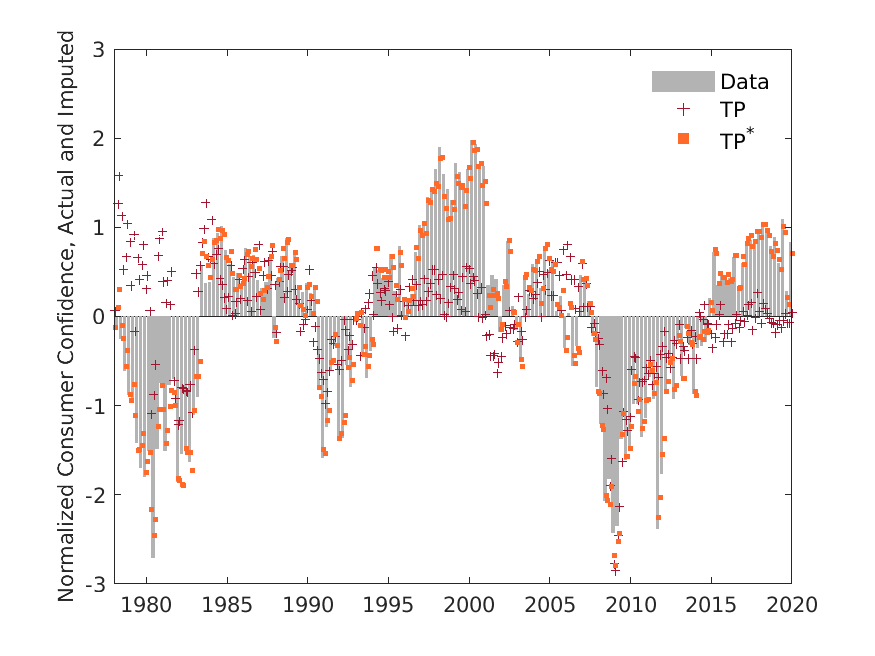}
\includegraphics[width=6.50in, height=3.5in]{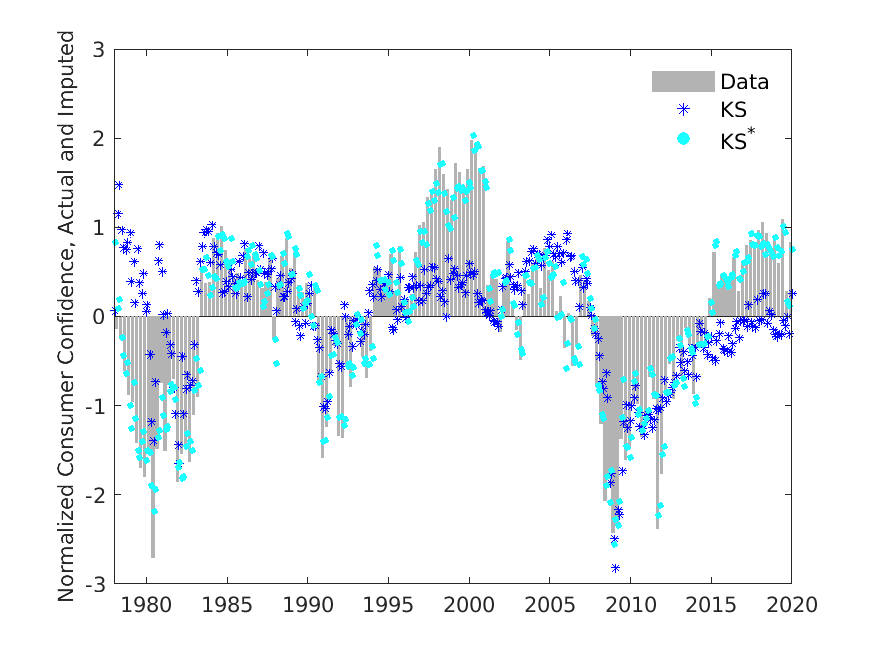}
The "Data" bars are the true monthly values of Consumer Confidence. In a hypothetical experiment, it is assumed that quarterly values are missing - the monthly values are imputed. The imputed monthly values using the different procedures are denoted by the markers. The * superscript estimators model serially correlated residuals. 
\end{figure}
\end{center}

We  now return to the  example considered earlier where $Y$ is the incompletely observed consumer sentiment series, and show that treatment of the dynamics of the idiosyncratic error is more important than the estimation procedufre for the common factors. The top  panel of Figure \ref{fig:cs-dynamic} compares the static \textsc{tp}  estimate  with   the dynamic estimate  \textsc{tp*}, where the latter includes one lag of both $F$ and  $Y$. 

We evaluate imputation errors  for the full sample and seven subsamples. \citet[Chapter 2.6]{vanbuuren:book} emphasizes that    imputation is not prediction and that the mean-squared error (MSE) is an uninformative  metric for evaluating imputation methods. The argument, made in the context of estimating slope parameters when the data are incompletely observed, is that the evaluation would treat the missing values as  random  and sampling error should be taken into account.  Our interest here is not estimation of the slope parameters but  the missing values themselves, and in particular, whether  they are centered around the true values. Though the variances of $\tilde Y_{H,s_t}$ are all downbiased because of omitting the variability of $\epsilon_{s_t}$, all procedures omit the same term.  

  \begin{table}[ht]
  \caption{Mean-Squared Errors in Imputing Consumer Confidence}
  \label{tbl:mmcs}
  \begin{center}
    \begin{tabular}{lllllllllll}
      \ start & end & TP & MLE-h & KS & TP$^*$ & MLE$^*$  & PC-GLS$^*$ & KS$^*$ & CL\\ \hline 
  Jan-1978 &   Dec-2019 &  15.622 & 15.187 & 14.479 &  6.771 &  7.062 &  7.109 &  7.229 &  6.498 \\
  Jan-1978 &   Dec-1983 &  10.317 & 10.144 &  9.428 &  2.766 &  3.003 &  2.939 &  2.913 &  3.271 \\
  Jan-1984 &   Jun-1994 &   4.483 &  4.267 &  4.226 &  3.111 &  3.209 &  3.204 &  3.463 &  3.165 \\
  Jul-1994 &   Dec-2000 &   7.174 &  6.990 &  6.707 &  1.841 &  1.906 &  1.987 &  2.128 &  2.152 \\
  Jan-2001 &   Dec-2006 &   3.849 &  3.630 &  3.531 &  3.215 &  3.439 &  3.342 &  3.515 &  2.702 \\
  Jan-2007 &   Dec-2010 &   3.433 &  3.290 &  3.013 &  2.300 &  2.482 &  2.586 &  2.283 &  1.973 \\
  Jan-2011 &   Dec-2019 &   6.279 &  6.056 &  6.030 &  3.081 &  2.990 &  3.134 &  3.113 &  2.385 \\
\hline
    \end{tabular}
  \end{center}
  {\footnotesize Notes:   (i) TP: (PC-F,static);  (ii) MLE-h (GLS-F, static), (iii) KS (KS),  (iv) TP*: (PC-F, dynamic), (v) MLE*: (gls-F,dynamic);   (vi)  PC-GLS* (GLS-F, dynamic); (vii)  KS* (KS, dynamic), (viii) CL (Chow-Lin, dynamic).}
  \end{table}

With this caveat in mind,  Table \ref{tbl:mmcs} reports the MSE for four static and four dynamic estimates  (with a suffix of `*'). The last column, 'CL', refers to  the Chow-Lin procedure which we treat as static.\footnote{\citet{proietti:06} shows that the Chow-Lin procedure can be cast in a  state space setup. The smoothed estimates can be treated as dynamic.}  Of these,  \textsc{ks}  and \textsc{ks*} explicitly model the dynamics of $F$, while \textsc{mle-h} only  accounts  for heteroskedasticity.   But the  static imputation errors are much larger than the dynamic ones. In the full sample, dynamic imputation reduces the error by half and the difference is particularly pronounced in the subsamples before 2000.  Consistent with the simulation results presented earlier, the gains from incorporating lagged information in $e_Y$ into the prediction model are much larger than using alternative  estimators for $F$ and $\Lambda$. To reinforce this point, the bottom panel of Figure \ref{fig:cs-dynamic} plots  \textsc{ks} (which is a filtered estimate) along with \textsc{ks*}  (which is a smoothed estimate). Both use (\ref{eq:fadl}) for imputation and differ only in the estimator for $F$. The two are very similar, suggesting that   the improvements are mainly due  to  information in lagged $Y$. In retrospect, this is not surprising   because $\rho$ in this application is estimated to be well above 0.8.

The monthly data used to estimate the factors  are seasonally adjusted. Seasonally unadjusted data are generally more volatile, but monthly seasonal variations are strictly periodic.  To anticipate our analysis of seasonally unadjusted weekly data to follow,  we  add some seasonally unadjusted series to the monthly panel and repeat  the counterfactual exercise.   The dynamic procedures yield imputed values that closer to the observed values than the static procedures. However, we find that  \textsc{tp*}, which does not directly model the idiosyncratic noise,  is less sensitive to seasonal variations in the unadjusted data.

\section{An Imputed Weekly CFNAI }

In this section, we take the low frequency indicator $y$ to be CFNAI, a monthly diffusion index published by the  Chicago Fed towards the end of each calendar month. The goal is to impute its weekly values $Y$.   To this end, weekly data are collected and detrended\footnote{We first-differenced the rail series in logs which does not leave a large spike, and linearly or quadratically detrend the other series.} for 13 real activity variables and 22 interest rates, exchange rates, and stock market  variables from (monday) 01-01-1990 to (monday)  05-31-2021. The financial variables are quite collinear, and we eventually used all 13 real activity variables, consumer credit, two interest rates,  a stock market index, and two money supply variables to form a 19 variable weekly panel with details given in the Appendix.

Analysis of weekly data can be challenging for several reasons. First,  the sampling frequency is not constant. There are six years in our sample that have 53 weeks (1990, 1996, 2001, 2007, 2012, 2018), and there is at least one month in every year with five mondays.  As the  number of mondays in a month  changes from month to month,  $m_t$ varies with $t$. The problem is not conceptually difficult, but does require special handling of the details. The second issue is that  the size of weekly panel is not only smaller than monthly panels like FRED-MD (McCracken and Ng 2016), but also less complete. In our case,   11 of the 19 variables  have some missing values, some from outlier adjustments  around  the financial crisis. There are also  'ragged-edges' once the sample is extended  to the latest data available. Since the weekly panel is already rather small in size, we use \textsc{tp} to construct a complete data panel $X_o$ for the purpose of factor estimation.   To focus on the  historical weekly values of CFNAI, we  exclude observations after COVID-19, giving $T=1566$ from 1990-01-01 to 2019-12-31.

The third issue is that  some  weekly series display strong seasonalities.      The panel of seasonally unadjusted  weekly data is better characterized by
\[ X_{o}=F\Lambda_o^{'} + \mathbb S +e_{X_o} \]
where $ \mathbb S$ is  a weekly seasonal  component, assumed to be uncorrelated with the non-seasonal series specific noise $e_{X_o}$ and with the weekly factors, $F$. A weekly series  can be  more variable than the corresponding monthly series because of  seasonality.  An ideal seasonal adjustment would be to first estimate  $ \mathbb S$  and  create a seasonally adjusted panel $X_{o}-\mathbb S$.  But as discussed in \citet{ng-worldcongress,guha-ng:22},   off the shelf seasonal adjustment procedures developed for monthly/quarterly data tend not to be adequate for  weekly data because the seasonal variations  are not strictly periodic.  The common approach, also taken in the construction of WEI and ECI, is   to remove seasonality by taking 52 week differences when needed. Though this removes a good deal of the seasonal variations, it also creates large  spikes because  Thanksgiving and Easter in particular   do not fall on the same week every year.  Systems approaches that jointly model the weekly and monthly data would ideally  also model these data irregularities, which is a daunting task.

\begin{center}
\begin{figure}[ht!]
\caption{Weekly Chicago Fed National Activity Index: Dynamic vs Static}
\label{fig:cfnai}
\includegraphics[width=6.5in,height=3.5in]{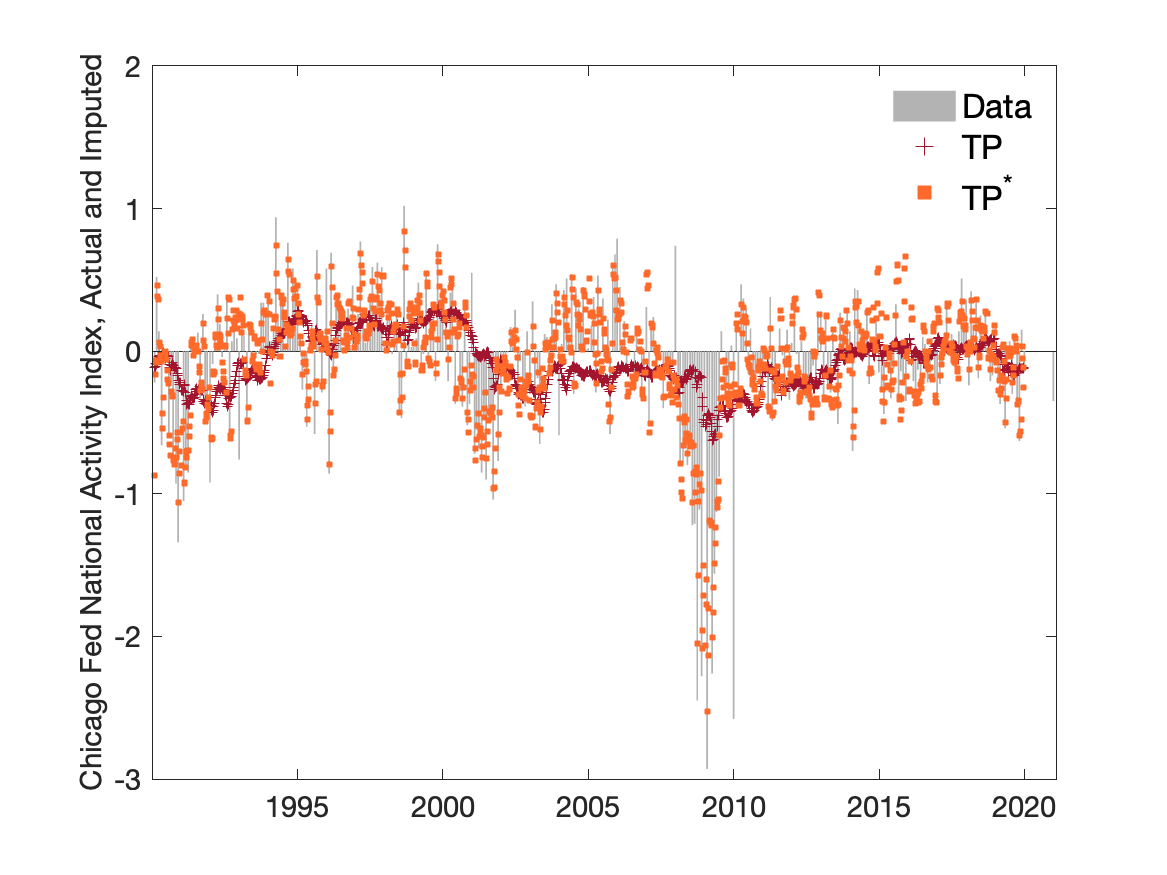}
\includegraphics[width=6.5in, height=3.5in]{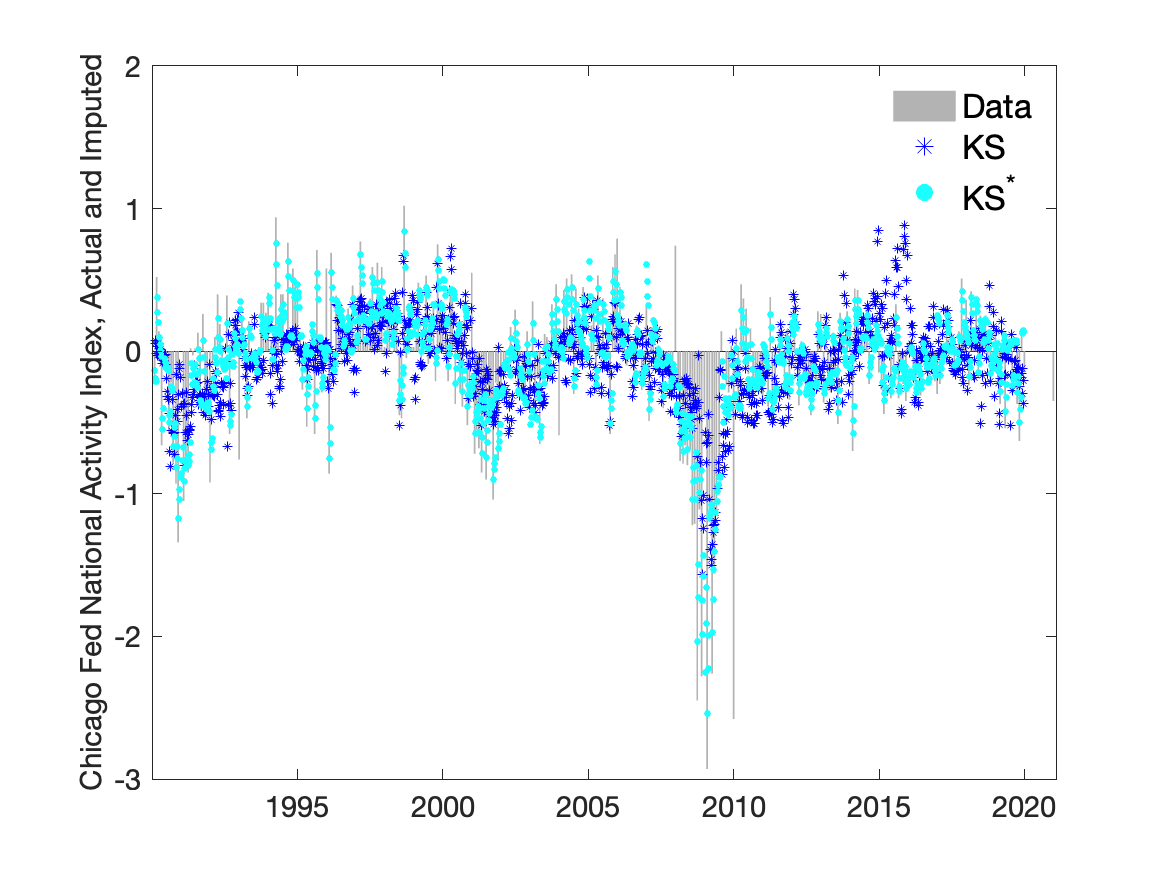}
The "Data" bars are the true monthly values of the Chicago Fed National Activity Index. Imputed "missing" weekly values using the different procedures are denoted by the markers. The * superscript estimators model serially correlated residuals.
\end{figure}
\end{center}

Our approach takes as starting point that so long as  seasonal variations are idiosyncratic and serially uncorrelated, they  play no role in imputation. We thus consolidate $ \mathbb S$ with the  idiosyncratic noise $e_{X_o}$ into a composite error. To smooth out the  seasonal variations on the factor estimates and to make better use of information available, we  expand    $X_o$ to include three lags of each variable, which also increases the size of the panel from 19 to 76 series. The three  factors in the expanded panel  account for 62\% of the variations in the data. The first weekly factor, which  explains 35\% of the variations in the data, loads most heavily on current and past values of unemployment claims.  The second factor loads on current and past consumer credit, while the third factor loads on current and past oil price.  Once the factors are estimated, we implement static or dynamic imputation. By construction, the imputed values in the last week of  each month  are the observed monthly  values. 

\begin{center}
  \begin{figure}[ht!]
  \caption{A Closer Look at Weekly Economic Activity 2011-2019} 
  \label{fig:zoom11-19}
  \includegraphics[width=6.5in,height=3.5in]{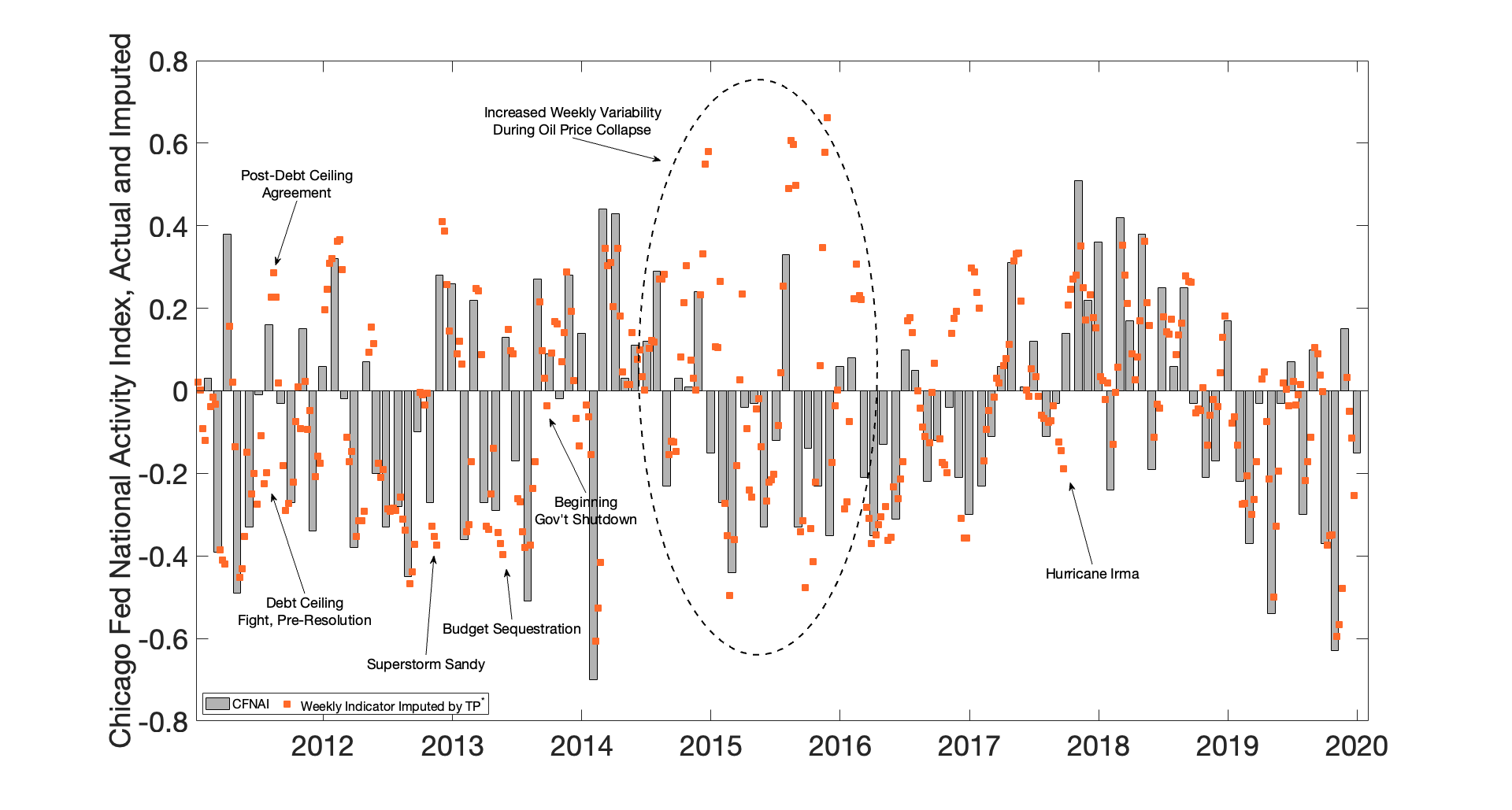}
  \end{figure}
\end{center}

The top panel of Figure \ref{fig:cfnai}  compares the static and dynamic \textsc{tp} procedures (\textsc{tp} and \textsc{tp*}, respectively), and the bottom panel compares the static and dynamic \textsc{ks} procedures.  The imputed values using the static \textsc{tp} model (dark red plus sign) is clustered closer to zero. The \textsc{ks} method shows only slightly more variation, and additionally appears to pick up some of the seasonal effects from the raw weekly data.  Accounting for persistence of idiosyncratic component $Y$ yield significant improvements, as shown by the \textsc{tp*} (orange square) and \textsc{ks*} (light blue circle) procedures. While \textsc{tp*}   only needs  PC estimation of the factors, \textsc{ks*} also need to model the dynamics of the factors and   the idiosyncratic noise.   Because the seasonal component is now consolidated into the noise, the error variance is very difficult to estimate.

We obtain seven imputed measures of CFNAI but will focus on \textsc{tp*} and \textsc{ks*} to illustrate the difference between dynamic and static imputation.  Our weekly imputed measures in general display lower volatility than the monthly measure, which is perhaps not surprising because the innovation at every $s_t$ is still set to the unconditional mean of zero. However, there are instances when the contribution of weekly information is notable.   The dynamic estimation procedures pick up weekly readings distinct from the monthly data around a number of short-lived events events such as hurricanes, ice storms, and the collapse of Long-Term Capital Management. Consider for example weekly values of CFNAI in the wake of the 2014-2015 collapse in oil prices.  As shown in Figure \ref{fig:zoom11-19}, the imputed weekly measures show greater variability than the monthly values during this time period.  Historically such an unexpected decline would be expected to provide a sizable boost to economic activity, but, as documented in \citet{baumeister-killian:16}, the boost to consumption was largely offset by the decline of investment in the oil sector. Contrary to previous episodes of oil price decline, the increased importance of shale production to the U.S. economy muddied the effect of the price drop.  The monthly series captures the net effects while the imputed weekly values preserve  these clashing forces.  A similar phenomenon is reflected during pre- and post-debt ceiling debates (2011 and 2013).

\section{Conclusion}
A by-product of static factor analysis in the presence of missing values is imputation of those missing values. These imputed values are unconditionally unbiased without modeling the persistence  of the idiosyncratic errors,  accounting for the serial correlation  provides  better estimates of the target than static  imputation. We have considered several ways by which residual serial correlation can be  incorporated into imputation. However, the user still has many modeling choices to make: how to estimate the factors, and whether to use parametric state space modeling  or a partial approach.   Dynamic imputation complicates inference, and  an assessment of the sampling uncertainty associated with the dynamic procedures  remains a topic for future research. This is studied in  \citet{goncalves-ng:23}  in the context of imputing  counterfactual outcomes.

\newpage
\section*{Appendix A: Estimators of Static Factors}
\paragraph{OLS and GLS Based Principal Components}
 The PC estimator for complete data   minimizes the objective function $\frac{1}{NT}\sum_{i=1}^N \sum_{t=1}^T (X_{it}-\Lambda_i'F_t)^2$ with respect to $F$ and $\Lambda$.\footnote{Assuming  $N$ is fixed,  maximum likelihood estimation gives consistent and asymptotically normal estimates of $\Lambda$ as $N\rightarrow \infty$ but not of  the factors.  Frequency domain methods can also consistently estimate the loadings in models with a dynamic structure,  as in \citet{geweke-77,sargent-sims,stock-watson:91}.}  Let $U$ and $V$ be  singular vectors in the singular value decomposition  $\frac{X}{\sqrt{NT}}=UDV'$,  where $D$ is a diagonal matrix of singular values arranged to be  in decreasing order.  Under the normalization that $F'F/T=I_r$ and $\Lambda'\Lambda$ is diagonal, the principal component (PC) estimates are defined by
\[ (\tilde F, \tilde \Lambda)=(\sqrt{T} U_r,\sqrt{N} V_rD_r)\]
where $U_r$ and $V_r$ are the left and right singular vectors corresponding to the $r$ largest singular values collected in the diagonal entries of $D_r$. The  PC estimates repeatedly perform  OLS  regressions of   $X_t=(X_{1t},\ldots, X_{Nt})'$ on the most recent estimates of $\Lambda=(\Lambda_1,\ldots,\Lambda_N)'$, and  of  $X_i=(X_{i1},\ldots, X_{iT})'$ on the most recent estimates of $F=(F_1,\ldots, F_T)'$. The  properties of $(\tilde F,\tilde\Lambda)$ are analyzed in \citet{stock-watson-jasa:02, baing-ecta:02, bai-ecta:03}, among others.\footnote{ \citet{fhlr-restat} represent the data as $x_{it}=\lambda_i(L)' g_t+e_{it}$ and estimate the $q$ dynamic factors $g_t$,  assumed to evolve as $g_{t}=C(L)u_t$, by   the method of dynamic principal components.}

Improvements to PC can be obtained even if $\Phi$, the misspecified idiosyncratic error variance, is not jointly estimated with $\Lambda$ as in maximum likelihood.   One  idea,   noted in \citet{boivin-ng-joe} and others,  is that PC  estimation   ignore the non-spherical error structure of the idiosyncratic errors and hence inefficient. Non-sphericalness  can  arise if the errors  are  serially correlated or heteroskedastic.    Suppose that $e_{it}$  is  a first order  autoregressive process, $ e_{it} =\rho_i e_{it-1}+\epsilon_{it}$,  $ \epsilon_{it}\sim (0,\sigma^2_{\epsilon_i}).$ If $F$ was observed, the efficient estimator of $\Lambda_i$ is a GLS regression of $X_i$ on $F$ which amounts to applying OLS to quasi-differenced  $X_{it}$  and  quasi-differened $ F_t$:
\begin{equation}
\label{eq:qd-lambda}  [(1-\rho_i L) X_{it}]= [ (1-\rho_i L)]   F_t']\Lambda_i+\text{error}.
\end{equation}
While controlling for serial correlation in  $e_{it}$ is useful in  estimation of  $\Lambda$,  accounting for  heteroskedasticity  by GLS is important for estimation of $F$.
\citet{breitung-tenhofen:11}  suggest to  update the PC estimate of $F$  using an estimate of $\Phi$. We will subsequently refer to  the PC estimator that  only accounts for heteroskedasticty as  PC-GLS-h.  If the $\Lambda$ estimates are also updated, it will be  referred to  as PC-GLS-har since it accounts for both heteroskedasticty and serial correlation.  See \citet[Section 7]{baili:12} and \citet{dgr-restat}.\footnote{ \citet{dgr-restat,dgr-joe} also considers quasi-maxiumum likelihood estimation of the factors as a function of the estimated loadings and variances.}

\paragraph{Quasi Maximum Likelihood Estimator}
As the name suggests, estimators based on quasi maximum likelihood directly maximize a misspecified likelihood function. \citet{baili:12,baili:16} jointly estimate $\theta=(\Phi,\Sigma_{F},\Lambda)$ 
by   maximizing the quasi-log-likelihood 
\[ \log L=\frac{1}{2N}\log |\Sigma_{X}(\theta)|-\frac{1}{2N} \text{tr} \bigg( S_{X}\Sigma_{X}^{-1}(\theta)\bigg)\] 
where $\Sigma_{X}(\theta)=\Lambda \Sigma_{F}\Lambda + \Phi$ is the covariance matrix implied by the model, $\Sigma_{F}$ is the $r\times r$  covariance of  $F$, and $S_{X}$ is the sample covariance of $X$. The misspecification is  that a diagonal $\Phi$ is assumed for $\Sigma_e$ even when it is not diagonal. The  estimator for $\Lambda$, say $\hat\Lambda$, solves the system of first order conditions with respect $\Lambda$, $\Phi=\text{diag}(\Sigma_e)$, and $\Sigma_{FF}$.
Assuming  joint normality of $X$ and $F$,  an  infeasible projections estimator for $F$ under this framework can be defined as:
\begin{equation*}
F_t^p=E_\Theta[F_t|X_1,\ldots,X_T]=\bigg(\Sigma^{-1}_{F}+ \Lambda'\Phi^{-1}  \Lambda\bigg)^{-1}  \Lambda' \Phi^{-1} X_t.
\end{equation*}
An alternative is the infeasible GLS estimator:
\begin{equation*}
 F^{GLS}_t=(\Lambda' \Phi^{-1}  \Lambda)^{-1}\Lambda' \Phi^{-1} X_t.
\end{equation*}  
The feasible analogs are obtained by plugging in MLE estimates for  $ \Lambda$ and $\Phi$ into (\ref{eq:gls-F}). When $\Phi$ only allows for heteroskedasticity, we will refer to it as MLE-h. \citet{baili:12,baili:16} show that the estimator is more efficient than PC.

\paragraph{Accounting for Dynamics in $F$.}
 The  static factors are identified in PC estimation by cross-section smoothing and ignores intertemporal information.  The QMLE framework can be used to exploit information available in the dynamics of $F$. Let  $\mathcal X=\vec(X)$  and  $\Sigma_{\mathcal F}$ be the matrix of autocovariances of the  $rT\times 1$ vector $\mathcal F=\vec(F')$. A projections estimator that takes the dynamics of $F$ into account  is 
 $F^{p}_t= \mathbf U_t'\mathcal F^p$ where $\mathbf U_t'= (\iota_t\otimes I_r) $, $ \iota_t$  is the $t$-th column of $I_T$.,
\begin{eqnarray*} \mathcal F^{p}= E_\Theta(\mathcal F|\mathcal X]&=&\bigg(\Sigma_{\mathcal F}^{-1}+(I_T\otimes \Lambda'\Phi^{-1}\Lambda)\bigg)^{-1}\bigg( (I_T\otimes \Lambda' \Phi)\bigg) \mathcal X.
\end{eqnarray*}
  Computation of  of $\Sigma_{\mathcal F}$  requires a dynamic model for  $\kfF$. If $\kfF$ is assumed to be a VAR of order $p$,   the state space representation  of $X$ is
\begin{align*}
  X_t &= \kfL \kfF_t + \epsilon_t && \epsilon_t \sim N\left(0,\Sigma_{\epsilon} \right) \\
  \kfF_{t+1} &= \kfVAR_1 \kfF_t + \kfVAR_2 \kfF_{t-1} + \dots + \kfVAR_p \kfF_{t-p+1}  + \eta_t && \eta_t \sim N\left(0,\Sigma_{\eta}\right)
\end{align*}

\citet{dgr-joe} use the  Kalman smoother to obtain   $\hat\Sigma_{\mathcal F}$ for an improved estimate of $F$. This, together with the OLS estimates $\hat\Sigma_e$  and $\hat\Lambda$ yields $\hat F_t=\mathbf U_t'\hat {\mathcal F}^p$. We will refer to this hybrid estimator  as PC-KS. State space models is fully parametric and  is efficient if the assumptions are correct. Section 5.2 suggests a way to  rewrite the observation and transition equations to  accommodate serially correlated  idiosyncratic error.

\clearpage

\section*{Data Appendix}
{\small
\begin{center}
\begin{tabular}{|l|l|} \hline
{\bf Variable} & {\bf Source} \\ \hline
Electric Utility Output & Edison Electric Institute, WRDS\tabularnewline
\hline 
Raw Steel Production & American Iron and Steel Institute, WRDS\tabularnewline
\hline 
Rail Traffic: Total & Association of American Railroads, WRDS\tabularnewline
\hline 
Rail Traffic: Class 1 Intermodal Units & Association of American Railroads, WRDS\tabularnewline
\hline 
Rotary Rig Count & Baker Hughes, WRDS\tabularnewline
\hline 
Unemployment Insurance Claims: Initial & U.S. Department of Labor, WRDS\tabularnewline
\hline 
Unemployment Insurance Claims: Cont. & U.S. Department of Labor, WRDS\tabularnewline
\hline 
Redbook, Same Store Sales & Redbook Research Inc., WRDS\tabularnewline
\hline 
WTI Oil Price & Commodity Research Bureau\tabularnewline
\hline 
CRB Commodity Price Index & Commodity Research Bureau\tabularnewline
\hline 
Mortgage Application Survey: Avg Total Loan & Mortgage Bankers Association\tabularnewline
\hline 
Mortgage Application Survey: Market Index & Mortgage Bankers Association\tabularnewline
\hline 
US Fuel Sales to End Users & U.S. Energy Information Administration\tabularnewline
\hline 
Consumer Credit & Board of Governors of FRS\tabularnewline
\hline 
Commercial Paper & Board of Governors of FRS\tabularnewline
\hline 
3-Month Treasury & Board of Governors of FRS\tabularnewline
\hline 
C\&I Loans, Large Banks / M1 & Board of Governors of FRS\tabularnewline
\hline 
S\&P Industrial / M1 & Standard and Poor's / BoG\tabularnewline
\hline 
S\&P 500 & Standard and Poor's\tabularnewline
\hline 
\end{tabular}
\end{center}
}

\clearpage

\bibliographystyle{harvard}
\bibliography{weekly,factor,bigdata,metrics,metrics2,macro,seasonal}

\includepdf{dataDescriptionWeek.pdf}

\bibliographystyle{harvard}
\bibliography{weekly,factor,bigdata,metrics,metrics2,macro,seasonal}

\end{document}